\documentclass{cimart}

\usepackage{longtable}

\title[How books tell a History of Statistics in Portugal]{How books tell a History of Statistics in Portugal: Works of foreigners, $\boldsymbol{estrangeirados}$, and others}

\author{% Please, use "Firstname Lastname" format, without abreviations
    Dinis Pestana and Rui Santos
    }

\authorinfo[% Format: please, use "F. Name"
    D. Pestana]{% Format: please, use "Department (only if 
    CEA/UL, FCUL, University of Lisbon, Portugal}{%
    ddpestana@ciencias.ulisboa.pt
    }

\authorinfo[% Format: please, use "S. Name"
    R. Santos]{% Format: please, use "Department (only if 
    CEA/UL, ESTG, Polytechnic Institute of Leiria, Portugal}{%
    rui.santos@ipleiria.pt
    }

\abstract{%
    Foreigners and ``estrangeirados'', an expression meaning ``people going to a foreign country [``estrangeiro''] getting there further education'', had a leading role in the development of Mathematical Statistics in Portugal. In what concerns Statistics, ``estrangeirados'' in the nineteenth century were mainly liberal intellectuals exiled for political reasons. From 1930 onwards, the research funding authority sent university professors abroad, and hired foreign researchers to stay in Portuguese institutions, and some of them were instrumental in the importation of new concepts and methods of inferential statistics. After 1970, there was a huge program of sending young researchers abroad for doctoral studies. At the same time, many new universities and polytechnic institutes have been created in Portugal. After that, aside from foreigners who choose to have a research career in those institutions and the ``estrangeirados'' who had returned and created programs of doctoral studies, others, who hadn't the opportunity of studying abroad, began to play a decisive role in the development of Statistics in Portugal. The publication of handbooks on Probability and Statistics, thesis and core papers in Portuguese scientific journals, and also of works for the layman, reveals how Statistics progressed from descriptive to a mathematical discipline used for inference in all fields of knowledge, from natural sciences to methodology of scientific research.
    }


\VOLUME{32}
\YEAR{2024}
\ISSUE{3}
\firstpage{485}
\DOI{https://doi.org/10.46298/cm.14005}

\begin{document}

\vspace*{1cm}
\begin{minipage}{0.03\textwidth}
	\begin{itemize}
		\item []
	\end{itemize}
\end{minipage}
\begin{minipage}{0.25\textwidth}
  \begin{itemize}
		\item []\includegraphics[width=\linewidth]{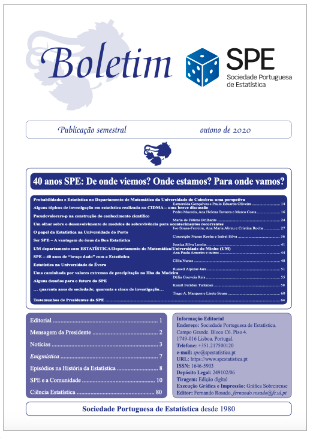}
	\end{itemize}
\end{minipage}
\begin{minipage}{0.40\textwidth}
	\begin{itemize}
		\item  [] Dedicated to Fernando Rosado, in gratitude for his outstanding  edition of the {\it Boletim  da Sociedade Portuguesa de Estat\'{i}stica}  and the {\it Memorial da Sociedade Portuguesa de Estat\'{i}stica}.
	\end{itemize}
\end{minipage} 
\begin{minipage}{0.25\textwidth}
  \begin{itemize}
		\item [] \includegraphics[width=\linewidth]{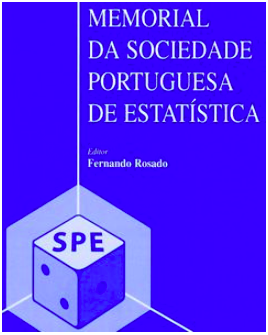}
	\end{itemize}
\end{minipage}

\begin{flushright}``Statistics is the grammar of science.''\\
Karl Pearson
\end{flushright}

\section{Introduction}

Sketching the History of Statistics in Portugal naturally requires discussing the influence of foreigners and ``$estrangeirados$",  foreign-influenced individuals\footnote{\, In the 17th and 18th centuries, some Portuguese intellectuals, who were exposed to scientific innovations and the Enlightenment philosophy abroad, had a far-reaching  influence on the modernization of culture in Portugal. Among these ``$Estrangeirados$'' (as they are referred to in Portuguese history) are Ribeiro Sanches, Luís António Verney, Francisco Xavier de Oliveira (Cavaleiro d'Oliveira), D. Luís da Cunha (whose political testament recommended Sebastião José de Carvalho e Melo, later the powerful prime minister  Marqu\^{e}s de Pombal,  to King D.\ José), and the Marquês de Pombal himself. The Pombaline reform of the university in 1772 was certainly due to the influential actions of the \textit{Estrangeirados}.

The liberal struggles of the 19th century forced many illustrious Portuguese to exile, particularly in the United Kingdom and/or France. Some of these new \textit{estrangeirados}, namely Solano Constâncio, Alexandre Herculano, and Oliveira Marreca, were important agents for the dissemination of Statistics in Portugal. The politician Passos Manuel, exiled in La Coruña-Plymouth-Belgium-Paris between 1828 and 1832, carried out important education reforms during his brief tenure as Minister of the Kingdom in 1836-1837, both in secondary and higher education. One of the consequences of  his reform was the creation of a course of Economia Pol\'{i}tica  [Political Economy], whose Professor,  Adri\~{a}o Forjaz de Sampaio, included the teaching of {\it Princ\'{i}pios de Statistica} [Statistical Primer]  \cite{Forjaz1841}  in its program from 1841 onwards.

We use the term $estrangeirado$ in a very general sense: \textit{estrangeirados} will also be much later  Zaluar Nunes (first as a student in Paris, then as a political exile in Brazil, due to the diaspora of intellectuals persecuted by Salazarism), Bento Murteira in London, Gustavo de Castro in Paris, Sebastião e Silva in Italy, Dias Agudo, Tiago de Oliveira, and Pedro Braumann in the USA, and later many scholarship holders who pursued doctorates abroad.}, without neglecting the role of   ``others''. Plagiating  Mae West, who    when asked what kind of men she preferred,    candidly replied: ``{\it Personally, I like two types of men --- domestic and foreign}", in our opinion many ``{\it Others}", who hadn't the opportunity of  extended partnerships abroad, had also a relevant role in the development of  Mathematical Statistics in Portugal, and their contributions   must be recognised and praised as much as those of the \textit{estrangeirados}. Notorious examples are Ant\'{o}nio Sim\~{o}es Neto, a man of culture and one of the early enthusiasts of Bayesian ideas in Portugal, who influenced many students at the Faculty of Sciences of Lisbon (namely Maria Ant\'{o}nia Amaral Turkman and Dinis Pestana), Fernando Rosado (to whom we dedicate this sketch of the history of Statistics in Portugal, thanking his contribution to  a ``{\it Portugaliae Monumenta Statistica}" with his fine edition of the {\it Boletim da Sociedade Portuguesa de Estatística} \cite{BolSPE} [Bulletin of the Portuguese Statistical Society]) and of the {\it Memorial da Sociedade Portuguesa de Estat\'{i}stica} \cite{MemSPE} [Memorial of the Portuguese Statistical Society], Salom\'{e} Cabral or Manuela Neves (recently recipient of the  Portuguese Statistical Society {\it Life Achievement Award}), whose doctoral dissertations were supervised by Tiago de Oliveira, or Rita Vasconcelos, who contributed so much for the progress  of the Probability and Statistics group at the University of Madeira. 

Regarding \textit{estrangeirados}, there is quite a bit of information, but when it comes to the influence of foreigners, the gaps are significant, especially considering the enormous long-distance influence of works already mentioned in the 19th century (Quételet by Forjaz de Sampaio, Bertrand and Poincaré in Sidónio Paes's thesis) or used as recommended bibliography in courses or research seminars (for example, books by Fisher, Kendall and Stuart, Cramér, Loève, Marek Fisz, Feller, Gnedenko, Cox, Zar, Rohatgi, ...).

Information is also quite lacking about the actual role of foreigners who came to Portugal as invited congress participants, and even more so about those who came on their own initiative as participants. There is also little information on the relevance to the development of Statistics in Portugal of those who served on advisory boards of research centers.

On the other hand, it is interesting to list which foreign statisticians were awarded honorary doctorates by Portuguese universities, which foreigners chose to spend part of their academic life in Portugal, and which books were translated and served as useful aids in the teaching, research, and dissemination of Statistics in Portugal.

Furthermore, research in university libraries and the National Library allowed us to identify and consult a reasonable number of statistical works that have been published, providing a reliable indicator of how the introduction of Mathematical Statistics concepts and the evolution of their teaching and dissemination progressed.

 Although we are aware that we are far from an exhaustive presentation, we will attempt to give a coordinated overview of how the teaching and dissemination of Statistics developed in the 19th century (Section 2). During this period, the increasing  importance of official statistics and the  introduction of  Probability  teaching at the Polytechnic School contributed to the recognition of Statistics as a mathematical discipline,  culminating with Sid\'{o}nio Paes   submission  of the first doctoral thesis in the field of  Mathematical Statistics in Portugal,  at the Faculty of Mathematics of the University of Coimbra. 

 Concerning the advancements in Mathematical Statistics in the 20th century up to 1974 (Section 3), we highlight   the consequences of starting  the teaching of Probability  at the Faculdade de Ciências [Faculty of Sciences].  Excellent professors developed the teaching of Statistics at  the  Instituto Superior de Agronomia (ISA) [Higher Institute of Agronomy] ---  Varennes e Mendon\c{c}a, Zaluar Nunes, and  Sebastião e Silva --- and at the Instituto Superior de Economia e Gest\~{a}o (ISEG) [Higher Institute of Economy and Management],  namely Leite Pinto and Bento Murteira, who worked under Maurice Kendall in London. These professors  imported many new statistical concepts, particularly the innovations created   by   K.\ Pearson, `Student', Neyman and E.\ Pearson, Fisher, and their co-workers.  Fisher's influence was significant at ISA, but we should not overlook the potential role the Institute of Anthropology in Coimbra could have played due to the hiring of Stevens, a disciple of Fisher, who published a series of remarkable ``Methodological Issues'' in the {\it Revista da Faculdade de Ciências da Universidade de Coimbra} [Review of the Faculty of Sciences of the University of Coimbra]. Finally, we include some informal notes  about the last fifty years (Section 4), for which it seems premature to adopt a historical perspective.

\section{19th Century}
The Venetian geographer and statistician Adriano Balbi (1782-1848) was in Portugal in 1820, before settling in Paris from 1821 to 1832, where he changed his name to Adrien Balbi. It was in Paris, in 1822, that he published the two volumes of the {\it Essai statistique sur le royaume de Portugal et d'Algarve}  \cite{Balbi1822a} [Statistical Essay on the Kingdom of Portugal and Algarve],  and also {\it Variétés politico-statistiques de la monarchie portugaise}  \cite{Balbi1822b} [Various Political-Statistical Observations on the Portuguese Monarchy].

\begin{figure}[h]
\centering
\begin{tabular}{ccc}
\includegraphics[scale=0.25]{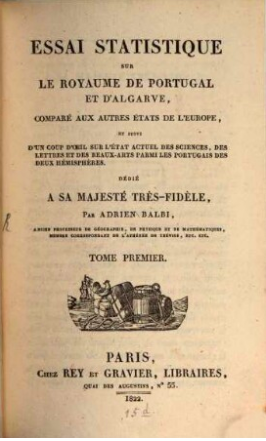} & 
\includegraphics[scale=0.18]{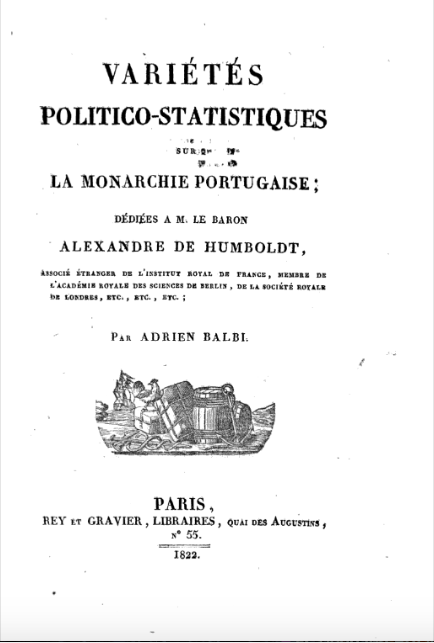} & \includegraphics[scale=0.4]{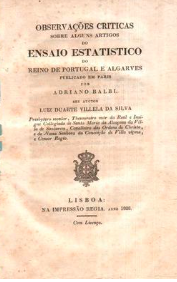}\\
(\textbf{a}) Essai Statistique& (\textbf{b}) Variétés Politico-Statistiques & (\textbf{c}) Critical remarks
\end{tabular}
\caption{Covers of Balbi's books on the Kingdom of Portugal and Villela da Silva's Critiques}
\label{SCbeta1}
\end{figure}
\medskip

Some statements in Balbi's books  probably irritated  the portuguese establishment, since in 1828 Villela da Silva \cite{Villela1828} published  at the Portuguese Royal Press negative critical observations on Balbi's works.

On the other hand, Balbi's work was  highly praised by Solano Constâncio (1777-1846), who in 1822 published an analysis of {\it Variétés politico-statistiques sur la monarchie portugaise} \cite{Solano1822} in the {\it Anais  das Ci\^{e}ncias, Artes, e Letras} [Annals of Sciences, Arts, and Letters], Volume XVII, 84--103.\footnote{\, Pages 212--222 of the 1995 reedition of {\it Leituras e Ensaios de Economia Política} \cite{Solano1995} [Readings and Essays in Political Economy], a selection of works by Solano Constâncio, published   by the Bank of Portugal, edited by José Luís Cardoso, accessible at 
%\href{https://www.bportugal.pt/sites/default/files/ocpep-11.pdf}{https://www.bportugal.pt/sites/default/files/ocpep-11.pdf}
\url{https://www.bportugal.pt/sites/default/files/ocpep-11.pdf}. The pages  cited ahead are from this edition.} It should be noted, incidentally, that Solano Constâncio, editor of the above mentioned  {\it Anais  das Ci\^{e}ncias, Artes, e Letras}, in this critical review refers to the article {\it Considera\c{c}\~{o}es sobre a Estat\'{i}stica} [Considerations on Statistics] by Cândido Xavier \cite{Xavier}, published in 1820, in Volume X, pp.\ 134--172.

Solano Constâncio was a remarkable expatriate polygraph, who studied Medicine in the United Kingdom between 1791 and 1797, later living in Paris from 1797 to 1799. He returned to Portugal, but in 1807 emigrated again to Paris. From 1822 to 1826 he lived in the United States of America, then returned to Paris, where he died in 1846. Machado de Sousa  \cite{Machado}  and  Cardoso  \cite{Cardoso}  are interesting sources of information about the wide range of interests of this remarkable intellectual.

Regarding Statistics, in the ``Preliminary Discourse'' \cite{Solano1818} of the  {\it Anais das Ciências, das Artes e das Letras}, Volume I (1818), pp.\ 1-37, (pp.\ 81-99 of {\it Leituras e Ensaios de Economia Política} \cite{Solano1995} [Readings and Essays in Political Economy]), Solano Constâncio states  that 
\begin{quotation}
``{\it Statistics, a science equally of our times, will also provide material for some articles and for examining the most notable works or news appearing on this subject.}" (p.\ 94)
\end{quotation}
and on p.\ 176 of the  {\it Anais das Ciências, das Artes e das Letras}, in the critical review of {\it An inquiry concerning population, etc..., ou investigação acerca da população e da faculdade de multiplicação da espécie humana; obra destinada a refutar a doutrina do ensaio de M. Malthus sobre esse assunto por W. Godwin, Londres, 1820} \cite{Solano1821}  [An inquiry concerning population, etc..., or an investigation into the population and the ability to multiply the human species; a work intended to refute the doctrine of Mr. Malthus's essay on this subject by W. Godwin, London, 1820] (Volume XII (1821), 63--104.):
\begin{quotation}
``{\it All the statistical notions that have been collected for a century agree in showing the great uniformity of nature's laws regarding reproduction of the species. The proportion of births and deaths, the number of males and females, the proportion of children per marriage, the number of marriages relative to the total population, and the age proportions, in historically regulated states, are facts that, the more they have been observed, the more conformity has been found between regions, climates, and diverse peoples. Despite the great imperfection of statistical works throughout Europe, those countries where censuses and population maps, birth and death registers have been pursued longer and with more care yield very analogous results among themselves. Among many other important inferences, it uniformly results from all these documents that the human species has a very small tendency to increase regularly and permanently.}"
\end{quotation}

The  {\it Armazém de Conhecimentos Úteis, nas Artes e Ofícios, ou Colecção de Tratados, Receitas e Invenções de Utilidade Geral} [Repository of Useful Knowledge, in Arts and Crafts, or Collection of Treatises, Recipes, and Inventions of General Utility], J.-P. Aillaud, Paris, 1838, contains ``Considerações sobre a estatística" \cite{Solano1838} [``Considerations on Statistics''] ({\bf 1}, 51--64), reproduced on pp. 271--277 of {\it  Leituras e Ensaios de Economia Política}\cite{Solano1995}  [Readings and Essays in Political Economy]; we quote from p.\ 276:
\begin{quotation}
``{\it  induction from statistical maps will prove the notable utility of this science, which is now indispensable for governing and enlightening the public.}"
\end{quotation}

We find it interesting to note that in the 1821 citation \cite{Solano1821} the refutation of Malthus'  exponential population growth, concluding that {\em ``the human species has very little tendency to increase regularly and permanently"} foreshadows what Verhulst \cite{Verhulst37,Verhulst45,Verhulst47} would later formalize, in 1837,  with the logistic model; and that, in the 1838 citation \cite{Solano1838}, {\em ``the notable utility of this science, which is now indispensable for {\rm [...]} enlightening the public"} has a meaning very close to the famous    Thomas Carlyle's   \cite{Carlyle} statement  in Chap. II of {\it Chartism}: {``\em A judicious man uses statistics, not to get knowledge, but to save himself from having ignorance foisted upon him.}".

More interesting still is the use in those excerpts of the words ``{\it inferences}'' and ``{\it induction}''. For Solano Constâncio, Statistics has an inferential component  that goes beyond its
 merely descriptive component. It is perhaps for this reason that he distinguishes ``{\it Estat\'{i}stica}"\  [Statistics]  from  ``{\it Estad\'{i}stica}'' [Stadistics],  etymologically derived from ``Estado" (government), the science of governing or dealing with political affairs, more connected to what is termed official statistics.

It was Francisco Solano Constâncio who, in 1836, in his {\it Novo Diccionario Critico e Etymologico da Lingua Portugueza} \cite{Solano1836} [New Critical and Etymological Dictionary of the Portuguese Language], defined, for the first time in Portuguese, Statistics  as the science ``{\it that deals with the enumeration of everything that forms the strength of a  nation}''.  He added a list of subjects: {\it population, agriculture, industry, education, public revenues and expenses, military strength, and property distribution}. The {\it Novo Diccionario Critico e Etymologico da Lingua Portugueza} achieved enduring editorial success, with 11 editions until 1877.

\begin{figure}[h]
\centering
\begin{tabular}{ccc}
\includegraphics[scale=0.20]{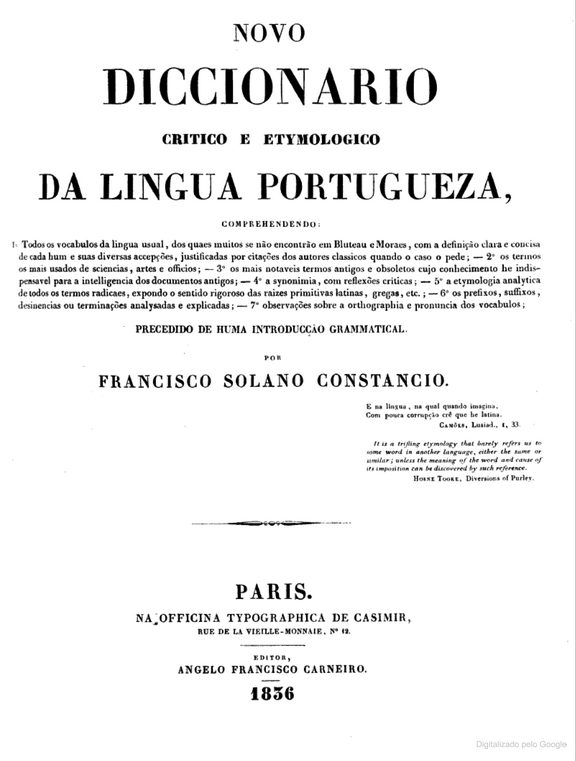} & 
\includegraphics[scale=0.60]{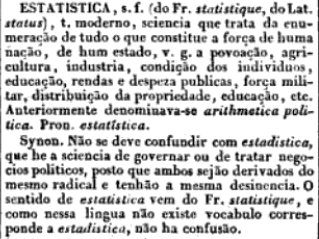} & \includegraphics[scale=0.22]{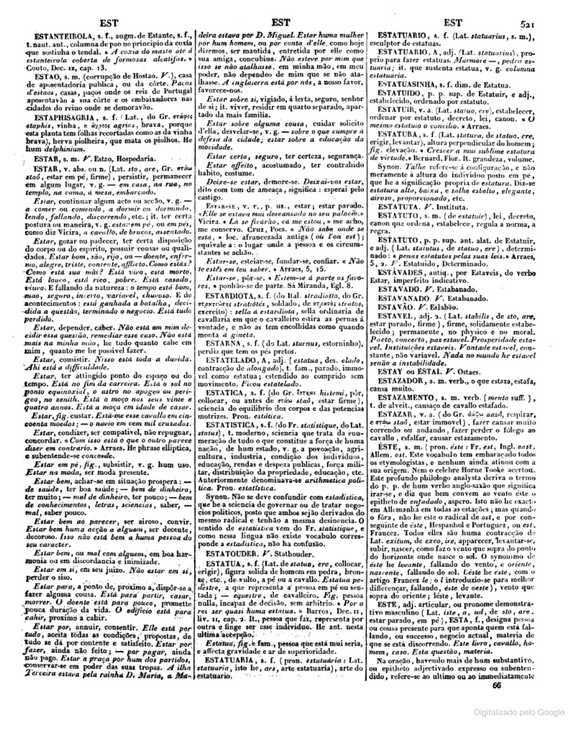}\\
(\textbf{a}) {\it Diccionario}'s cover& (\textbf{b})  entry ``ESTATISTICA'' & (\textbf{c}) page 66 of the {\it Diccionario}
\end{tabular}
\caption{Solano Const\^{a}ncio' s {\it Novo Diccionario Critico e Etymologico da Lingua Portugueza}}
\label{SCbeta2}
\end{figure}

One of the consequences of the 1836 education reform by Passos Manuel was the creation of a Chair of Political Economy at the Faculty of Law (resulting from merging the faculties of {\it Laws} and of {\it Canons}) of the University of Coimbra. Adrião Forjaz de Sampaio included in the program of the chair an informal presentation of Descriptive Statistics, from the perspective of preparing official statistics, heavily influenced by the journal of works of the Société Française de Statistique Universelle (Fontes de Sousa,  2005 \cite{Fontes}; 
 Ferrão,  2006 \cite{Ferrao}).

\noindent\begin{minipage}{0.25\textwidth}
  \begin{itemize}
 \item []
\includegraphics[width=\linewidth]{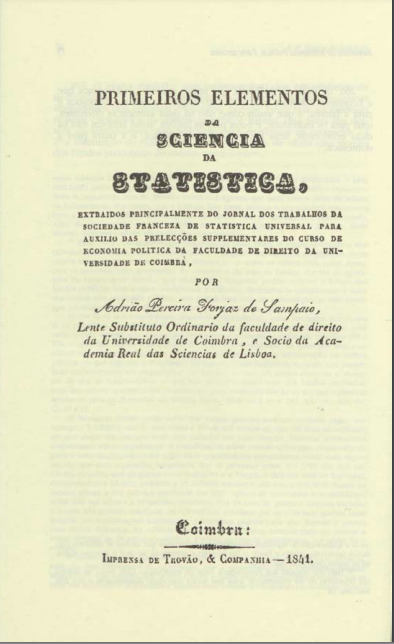}
\end{itemize}
\end{minipage}%
\hfill%
\begin{minipage}{0.75\textwidth}
\begin{itemize}
\item  [] 
Thus, the pioneer of Statistics teaching  at the Faculty of Law of the University of Coimbra was Adrião Forjaz de Sampaio (1810-1874). He published {\it Primeiros Elementos da Sciencia da Statistica}  \cite{Forjaz1841}  [First Elements of the Science of Statistics], in which the influence of Quételet is noted. This manual presents the  elements of Descriptive Statistics, giving some prominence to averages and dispersions, with rudimentary mathematical formalism. It is interesting to note that in the title, in later editions Forjaz de Sampaio uses ``Stadistics''\ instead of ``Statistics'', and that in 1857 José Dias Ferreira (1837-1907), then still a student, published at the University Press an {\it  Ensaio sobre os Primeiros Elementos da Theoria da Estadística do Exmo.\ Senhor Adrião Pereira Forjaz de Sampaio} \cite{Ferreira1857} [Essay on the First Elements of the Theory of  Stadistics  by the Honorable Adrião Pereira Forjaz de Sampaio].\end{itemize}
\end{minipage}\\
\bigskip

In 1887 José Frederico Laranjo \cite{Laranjo1887} published a new program for the Chair of Political Economy and Statistics at the Faculty of Law of the University of Coimbra, ``with the approval of Dr. Manuel Nunes Giraldes''. Citing Bastien    \cite{Bastien}  ``{\it Note that the Chair of Political Economy, after its creation in 1836, and until 1871, was mostly held by Adrião Forjaz de Sampaio and after that date, and until 1899, by Manuel Nunes Giraldes. Despite the strong relationship of Frederico Laranjo with the Chair of Political Economy, particularly in terms of teaching, program development, publication of teaching manuals, and doctoral thesis examinations [...]\ he was never awarded the respective chair.}''. The lessons of J.\ F.\ Laranjo in {\it Princípios de Economia Política} (1891) [Principles of Political Economy], reprinted in \cite{Laranjo}, do not contain notes on Statistics.
\bigskip

The dichotomy between {\it Estat\'{i}stica/Estad\'{i}stica}  [Statistics/Stadistics] continued to divide those interested in this new science(s) in the 19th century. On March 26, 1852, Alexandre Herculano\footnote{\, A remarkable historian, author of an excellent four volumes very comprehensive {\it Hist\'{o}ria de Portugal} [History of Portugal] and of {\it Hist\'{o}ria da Origem e Estabelecimento da Inquisi\c{c}\~{a}o em Portugal.} [History of the Origin and Establishment of the Inquisition in Portugal]. He  compiled and edited   {\it Portugaliae Monumenta Historica},  and was also a  highly praised romantic novelist author namely of historical  novels.} (1810-1877) --- who was exiled in England and then in France between 1831 and 1832 --- proposed to the  Lisbon  {\it Royal Academy of Sciences}   that the section of Economic and Administrative Sciences should   ``{\it draft the necessary instructions and a series of statistical questions, in harmony with the current state of science, which the Academy should bring to the Government's attention, seeking to obtain from it an order for civil, ecclesiastical, and military officials, as well as judicial magistrates and elected authorities of any order and denomination, to respond within the scope of their respective actions to the aforementioned questions, with these responses being transmitted to the Academy.}" 

This interest of Herculano in Statistics was not an isolated episode in his vast array of interests and activities. Two excerpts from {\it Emigra\c{c}\~{a}o} [{\it Emigration}], in {\it Opúsculos}, Volume IV \cite{Herculano}  show that Herculano maintained his interest in Statistics, appreciating its contribution to the objective understanding of social phenomena:

\begin{quotation}
{\it We thus have a set of respectable opinions on the insufficiency of rural wages. These opinions are fully confirmed by statistics, revealing with the irresistible eloquence of numbers the true situation of the laborer, both in terms of his resources and his needs. Incomplete for covering only part of the districts of the kingdom, deficient due to omissions and lack of specification in the elements provided by some municipalities, the statistical tables added to the Parliamentary Inquiry are still numerous enough in their various types to allow general conclusions to be drawn from them.} 
\flushright Alexandre Herculano, {\it Emigration}, III  
in {\it Opúsculos}, Volume IV, pp. 150-151.
\end{quotation}

\begin{quotation}
{\it If the rejection of official means of information seems inconvenient and unfair to me, I find the absolute condemnation of statistics itself, voted to ostracism as a conjectural science, even more peculiar. There is necessarily a misunderstanding here. Statistics aims to collect and methodically organize social facts that can be expressed in numbers. Nothing less conjectural. If, instead of facts, assumptions are collected, that is not statistics, it is fiction. In applied statistics, conjectures and the errors often derived from them are possible; but to infer from this the futility of statistics is the same as denying the scientific validity of arithmetic because sums or multiplications are often mistaken. Generally, what do the laws and measures of a generic nature, whether of a legal, moral, or economic order, depend on? They depend on a correlative statistical study. It is in this study that their raison d'être lies.}
\flushright Alexandre Herculano, {\it Emigration} VII (October 1874)\\
in {\it Opúsculos}, vol. IV, pp. 212-213.
\end{quotation}
\bigskip

António de Oliveira Marreca (1805-1889) also went into exile in England and France, returning to Portugal in 1833. He was highly influential in the institutionalization of Political Economy in Portugal, for   detailed information cf.\  Veríssimo Serrão       \cite{Serrao} and  Almodôvar    \cite{Almodovar}.
\medskip

\noindent\begin{minipage}{0.25\textwidth}
  \begin{itemize}
 \item []
\includegraphics[width=\linewidth]{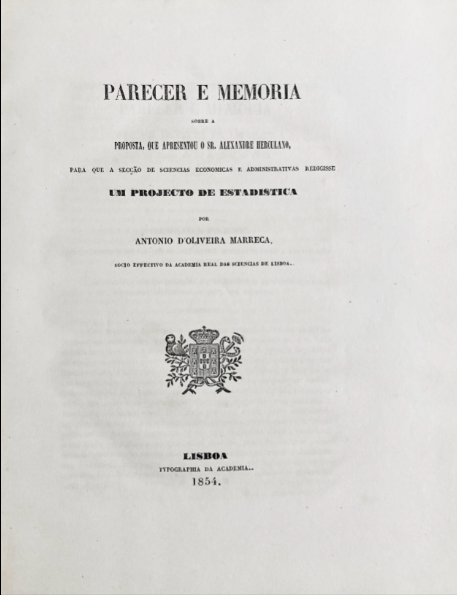}
\end{itemize}
\end{minipage}%
\begin{minipage}{0.7\textwidth}
\begin{itemize}
\item  []  In {\it Parecer e Memoria sobre a proposta, que apresentou o Sr. Alexandre Herculano: para que a secção de sciencias económicas e administrativas redigisse um Projecto de Estadística} \cite{Marreca1854} [A Review  and  Memoir on Mr. Alexandre Herculano proposal, requiring  that  the section of economic and administrative  sciences establishes a Project  on Stadistics],   Oliveira Marreca returns  to the use of the  terminology  {\it Estad\'{i}stica}. 
\end{itemize}
\end{minipage}
\bigskip

Oliveira Marreca was one of the administrators of the Imprensa Nacional de Lisboa [National Press of Lisbon]  from 1834 to 1838. Arguably Oliveira Marreca  influenced the publication at the Imprensa Nacional de Lisboa  of   the translation {\it Compendio Estadistico ou Resumo dos Elementos de Estadistica} \cite{Jonnes} [A Treatise on Statistics, or A Resume of the Elements of  Statistics]  of the 1847 {\it  \'{E}l\'{e}ments de Statistique: comprenant les principes g\'{e}n\'{e}raux de cette science, et un aper\c{c}u historique de ses progr\`{e}s } \cite{Jonnes1847} by A. Moreau de  Jonnès  (then in charge of the bureau of {\it Statistiques G\'{e}n\'{e}rales}  of France). It is worth noting that \, Jonnès clearly expresses his understanding of the scope and limitations of Statistics:  
\begin{quotation}
{\it Statistics is the science of natural, social, and political facts expressed in numbers. Its aim is the deep knowledge of society considered in its nature, elements, economy, situation, and changes. It uses the language of numbers, which is as essential to it as figures are to geometry and symbols are to algebra. It constantly proceeds through numbers, which gives it the precision and certainty characteristic of the exact sciences. Works that bear its name, without having its object or language, do not belong to it, as they fall outside the conditions of its existence. Thus, statistics without numbers, or whose numbers do not express natural, social, and political facts, do not deserve the name they bear. The same can be said of moral and intellectual statistics, because it is a ridiculous claim to want to subject the soul and passions to calculation, and to compute, as if they were defined and comparable units, the movements of the spirit and the phenomena of human intelligence.}
\end{quotation}
For detailed information on Jonnès and his influence, see Paixão Santos   \cite{Santos}.

The politician António José de \'{A}vila (1807-1881) --- the presumed inspiration of  Eça de Queiroz's (1925) posthumous satyrical novel {\it  O Conde de Abranhos}\footnote{\'{A}vila prohibited a series of liberal conferences ({\it Confer\^{e}ncias do Casino}), and the outstanding novelist E\c{c}a de Queiroz used him as model for the ridiculous politician    Conde de Abranhos in a satyrical novel, published only 25 years after the death of E\c{c}a de Queiroz.  As challenging of the political   establishment as {\it Yes, Minister} and {\it Yes, Prime Minister}.} --- also reinforced Alexandre Herculano's proposals in 1854 in a report he prepared for the Minister of Public Works Fontes Pereira de Melo on the works of the Statistical Congress held in Brussels in 1853. He emphasized the importance of establishing uniform bases for national statistical systems, highlighting their importance for the establishment of international comparisons.   Cardoso  \cite{Cardoso12} claims  that \'{A}vila  ``{\it supports the need for a general population census every 10 years, with the collection of basic identification and social status information (profession, education, social and health conditions, etc.). Among the subjects that should be the object of statistical standardization, he gives special attention to the following: territory and property registry; emigration and migratory movements; agricultural production; industrial statistics (production, employment, wages, capital, etc.); commercial statistics; budgets of the working classes; census of the indigent (poverty and social assistance); levels of education; criminality and repression.}"
 \'{A}vila  was again the portuguese official  delegate at  the Statistical Congress held in Berlin in 1863 \cite{Avila1864}.

There was also a ``folkloric'' aspect to the use of Statistics: Sexual taboos were also statistically addressed (Santos Cruz \cite{Santos1841}; Azevedo \cite{Azevedo1864}); and perhaps in an attempt to finance their expenses, some travellers published accounts of their peripatetic wanderings with some pretension of statistical objectivity, such as {\it Portugal a Voo de Pássara} \cite{Rattazzi1880}\footnote{\,  As  the novelist Camilo Castelo Branco insultingly translated the title {\it Le Portugal à Vol d'Oiseau} [{\it The Flight of a Female Bird in Portugal},] by Marie Rattazzi, A. Degorce-Cadot, 1880.  ``P\'{a}ssara" means literally ``female bird", but  its slang meaning is ``cunning", ``smart-ass", ``shrewed-ass", or even worse.}. In the posthumous publication {\it Londres Maravilhosa} [Wonderful London], in the Notes -- Excerpts from the notebooks of M. Teixeira-Gomes\footnote{\, Aside being a diplomat and President of the Portuguese Republic, Manuel Teixeira-Gomes  was a  notorious writer, author namely of the novel {\it Maria Adelaide}, published in 1938, describing the attraction for ``nymphets"\ earlier than the famous {\it Lolita} of Vladimir Nabokov. Fialho de Almeida was also an important 19th century portuguese novelist, obviously without any serious background in Statistics.  In 1905 he published {\it Cadernos de Viagem. Galiza} [{\it Travel Notebooks. Galiza}].}, he reveals, with a mixture of surprise and irony, that  the novelist and journalist  Fialho de Almeida  ``{\it is preparing for his excursion to Galiza; he plans to make a complete book with statistics and agronomy about Galiza!}" 

The teaching of elementary  Probability Calculus, included in the program of Infinitesimal Calculus,   was introduced  in 1865 at the Escola Polit\'{e}cnica de Lisboa [Lisbon Polytechnic School],  formerly Col\'{e}gio dos Nobres [Nobleman College] and later, in 1911, Faculdade de  Ci\^{e}ncias  [Faculty of  Sciences]. Statistics gradually detached itself from its subordination to Political Economy, and in the last quarter of the 19th century, the first research works linked to Mathematics   were published in  1866 and  1870 (Daniel Augusto da Silva \cite{Silva1866, Silva1870}) and later, in 1898, the first doctoral thesis on Mathematical Statistics (Sidónio Paes \cite{Paes1898}). 

The third posthumous edition (1756) of {\it The Doctrine of Chances} by Abraham de Moivre \cite{Moivre1756}  contains applications of Probability Theory to actuarial calculations regarding the computation of annuities. Daniel Augusto da Silva addressed the calculation of annuities in 1866, thus becoming the patron of Portuguese actuaries. It is an interesting investigation but of little statistical interest, the reference to   statistics is limited to the excerpt we reproduce.
  
\bigskip

\hspace*{-1cm} 
\begin{minipage}{0.22\textwidth}
  \begin{itemize}
 \item []
\includegraphics[width=\linewidth]{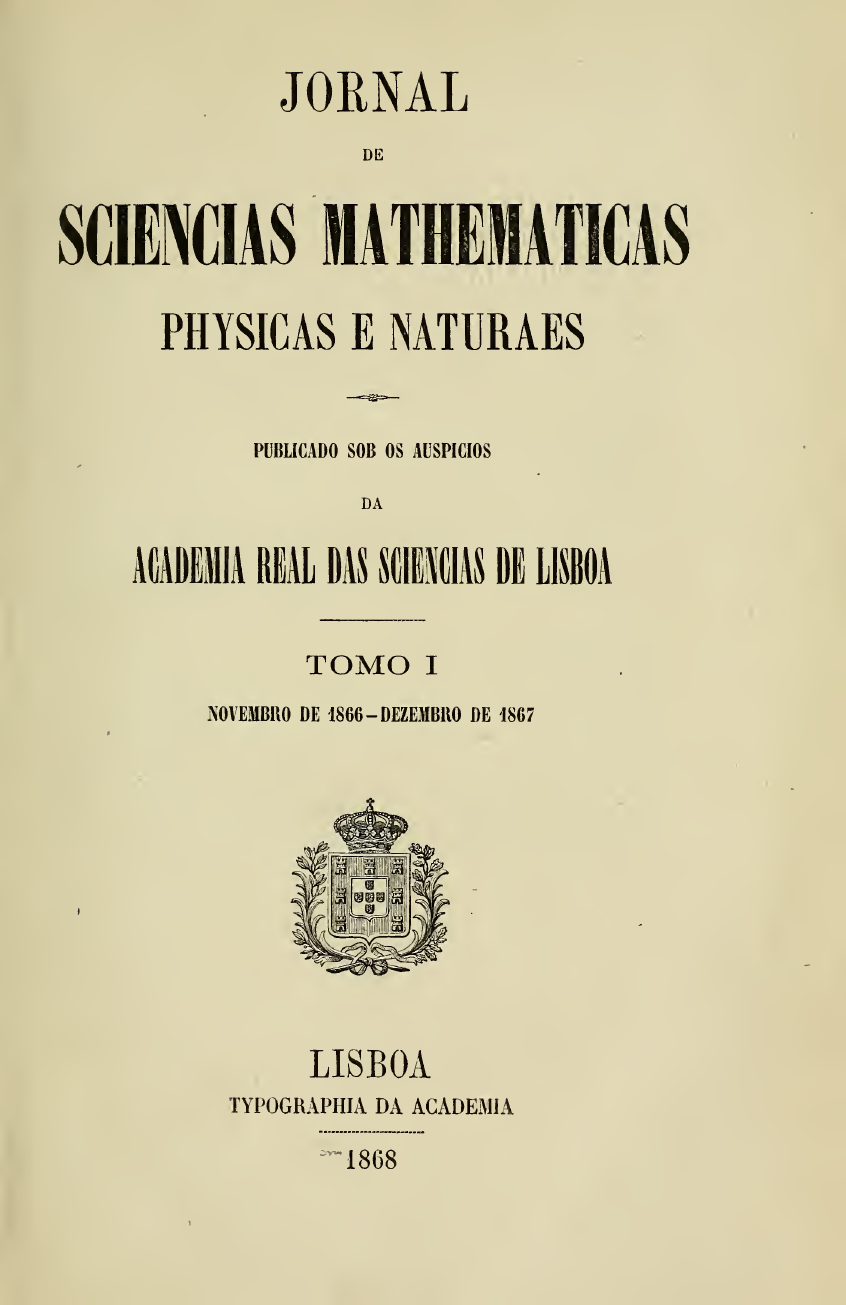}
\end{itemize}
\end{minipage}%
\begin{minipage}{0.5\textwidth}
\begin{itemize}
\item  []  
\includegraphics[scale=0.38]{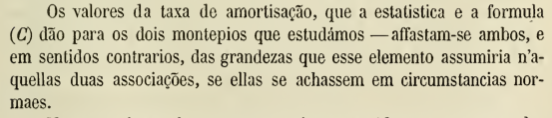}
\footnotesize {\it The values of the amortization taxes given by Statistics      and formula (C) for the two Montepio Mutualist Associations that  we analyse fall apart, in opposite directions,   from the values they would assume in those associations, in case they would be in normal circumstances.} 
\end{itemize}
\end{minipage}%
\begin{minipage}{0.25\textwidth}
  \begin{itemize}
 \item []
\includegraphics[width=\linewidth]{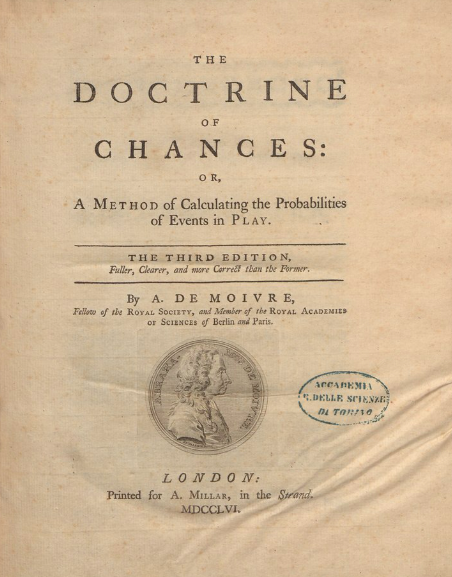}
\end{itemize}
\end{minipage}%
\bigskip

 Daniel Augusto da Silva  also   published a comparative study of population movements in Portugal \cite{Silva1870}. Detailed information can be found in Martins   \cite{Martins12}, with further complements in  Martins   \cite{Martins20}.  
\bigskip

\noindent
\begin{minipage}{0.7\textwidth}
\begin{itemize}
\item  []  The  {\it Geographia e Estatística Geral de Portugal e Colónias} \cite{Pery1875} [Geography and General Statistics of Portugal and the Colonies] by Gerardo A. Pery, from 1875,   is an interesting synthesis that, without methodological innovations, plays a pioneering role in the association of Geography with Statistics. Population is also a major theme in this work, as far as statistics is concerned.
\end{itemize}
\end{minipage}
\begin{minipage}{0.18\textwidth}
  \begin{itemize}
 \item []
\includegraphics[width=\linewidth]{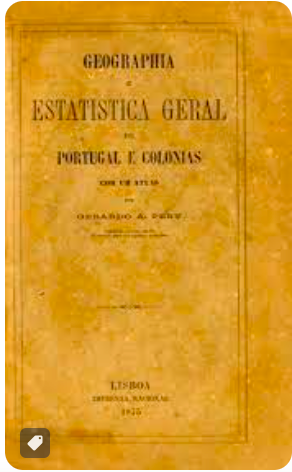}
\end{itemize}
\end{minipage}%
\bigskip

\noindent\begin{minipage}{0.2\textwidth}
  \begin{itemize}
 \item []
\includegraphics[width=\linewidth]{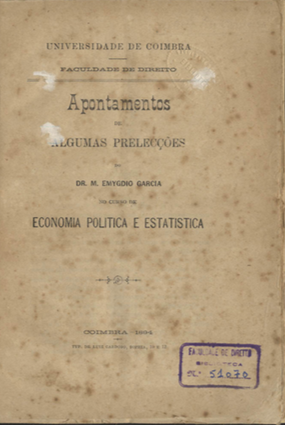}
\end{itemize}
\end{minipage}%
\begin{minipage}{0.7\textwidth}
\begin{itemize}
\item  []  From 1871 to 1899 Manuel Nunes Giraldes was the lecturer in Economics and Statistics at the Faculty of Law in Coimbra; in addition to José Frederico Laranjo's collaboration, M.\ Emydgio Garcia gave some lectures on the course. Emydgio Garcia published the corresponding notes in 1894 \cite{Garcia1894}, where Statistics was confined to the title of the pamphlet and a set of general comments, which seems to indicate some backlashes in the teaching of Political Economy at the Faculty of Law.   \end{itemize}
\end{minipage}
\bigskip

 Chapter IV, entitled {\it General Notion of Statistics}, pp.\ 14-17, is merely a set of debatable aphorisms, for instance  ``{\it Statistics is generally considered the (concrete and descriptive) science that collects, gathers, and records phenomena or facts of observation, expressing them in numbers and geometric signs. 
 
It is not properly a science; it is a necessary aid, an auxiliary, and moreover, an indispensable factor of science.}"

It is worth mentioning that ``mathematical'' Statistics had a slow start, mainly because it depends on the language of Probability. The works of John Graunt (1662) \cite{Graunt} and   Halley(1693) \cite{Halley} on mortality are pointed out as the origin of Demography and, indirectly, Actuarial Science, with the referred works of Daniel Augusto da Silva being late descendants.

Other pioneering works in Mathematical Statistics, notably the posthumous publication where Thomas Bayes  (1763) \cite{Bayes}   used Probability as the language of Statistics for the first time,   the great memoir of Laplace's  (1774) \cite{Laplace} on inference, and Gauss' research (1809) \cite{Gauss} in which the method of least squares is developed, only came to have an impact on statistical production in Portugal at the end of the 19th century.
\bigskip

 \hspace*{-1cm}
\begin{minipage}{0.75\textwidth}
\begin{itemize}
\item  []  The first doctoral thesis at the Faculty of Mathematics of the University of Coimbra in the field of  Mathematical Statistics, {\it Introdu\c{c}\~{a}o \`{a} Teoria dos Erros das Observações} \cite{Paes1898} [Introduction to the  Theory of Observational Errors], was defended in 1898 by Sidónio Paes. It proposes a rigorous demonstration of the approximation of sums by Gauss's law, and of the use of ``lesser squares''.  It explicitly mentions the works of Bayes, Laplace, Lagrange and Gauss, but without precise referencing. 
\end{itemize}
\end{minipage}
\begin{minipage}{0.25\textwidth}
  \begin{itemize}
 \item []
\includegraphics[width=\linewidth]{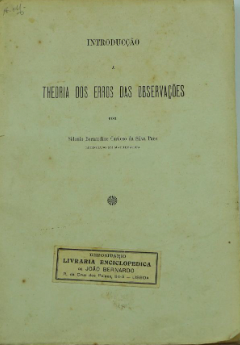}
\end{itemize}
\end{minipage}%
\bigskip
\vspace{0.35pc}

In fact, in Sid\'{o}nio Paes thesis, only the books by Poincaré and Bertrand on Probability Calculus, a work by Chauvenet and another by Estienne are identified in footnotes. Works by Encke, Schiaparelli, Stone, Ellis, Tait, Glaisher, Hagen, de Morgan, Ferrero, Bessel and Crofton are mentioned in the text, but without any precise indication.

At the end of the century, 
Pedro Afonso André de Figueiredo, honoured with the title of Viscount of Wildik, published {\it Notice Statistique sur le Portugal} \cite{Wildik1900} for the Universal Exhibition of 1900.
\bigskip

\subsubsection*{Chronology}
(English translation of Portuguese titles inside square brackets. Indication of libraries where rare books were found inside angle brackets: $<FLUL>$ -- Faculdade  de Letras da Universidade de Lisboa; $<FDUL>$ -- Faculdade  de Direito da Universidade de Lisboa; $<FMUL>$ -- Faculdade  de Medicina  da Universidade de Lisboa; $<IICT>$ -- Instituto de Investiga\c{c}\~{a}o Cient\'{i}fica Tropical; $<IST>$ -- Instituto Superior T\'{e}cnico; $<ISA>$ -- Instituto Superior de Agronomia; $<FPCEUL>$ -- Faculdade  de Psicologia e Ci\^{e}ncias da Educa\c{c}\~{a}o  da Universidade de Lisboa; $<FCUL>$ -- Faculdade  de Ci\^{e}ncias da Universidade de Lisboa; $<ISCSP>$ -- Instituto Superior  de  Ci\^{e}ncias   Sociais e Pol\'{i}ticas; $<ISEG>$ -- Instituto Superior de   Economia e Gest\~{a}o.)
\medskip

\noindent  {\bf 1818} --- F.\ Solano Constâncio's  [Preliminary Speech] \cite{Solano1818}.
\vspace{0.35pc}

\noindent  {\bf 1820} --- C.\ Xavier's [Considerations on Statistics] \cite{Xavier1820}.
\vspace{0.35pc}

\noindent {\bf 1821} --- F.\ Solano Constâncio's [An inquiry concerning population, etc..., or an investigation into the population and the ability to multiply the human species; a work intended to refute the doctrine of Mr.\  Malthus's  essay on this subject by W. Godwin, London, 1820], \cite{Solano1821}.
 \vspace{0.35pc}
 
\noindent {\bf 1822} ---  A.\ Balbi's 
{\it Essai statistique sur le Royaume de Portugal et d'Algarve, comparé aux autres états de l'Europe, et suivi d'un coup'oeil sur l'\'{E}tat actuel des sciences, des lettres et des beaux-arts parmi les portugais des deux hémishéres} \cite{Balbi1822a}. 
\vspace{0.35pc}

---  A.\ Balbi's 
{\it Variétés politiques et statistiques de la monarchie portugaise} \cite{Balbi1822b}.
\vspace{0.35pc}

 --- F.\ Solano Constâncio's [Variétés politico-statistiques sur la monarchie portugaise, by A. Balbi] \cite{Solano1822}.
\vspace{0.35pc}

\noindent {\bf 1828} --- L.\ D.\ Villela da Silva's [{\it  Critical observations on some considerations in Variétés politiques et statistiques de la monarchie portugaise, published in Paris by Adran Balbi}] \cite{Villela1828}. $<FLUL>$
\vspace{0.35pc}

\noindent {\bf 1836} --- F.\ Solano Constâncio's [{\it  New Critical and Etymological Dictionary of the Portuguese Language}] \cite{Solano1836}.
 \vspace{0.35pc}
 
---  Passos Manuel's  reform of the University, namely    creating of a Chair of Political Economy at the Faculty of Law of the University of Coimbra.
  \vspace{0.35pc}

\noindent {\bf 1838} --- F.\ Solano Constâncio's  [Considerations on Statistics] \cite{Solano1838}.
 \vspace{0.35pc}

\noindent {\bf 1841} --- A.\ P.\ Forjaz de Sampaio's [{\it  First Elements of the Science of Statistics}] \cite{Forjaz1841}. \vspace{0.35pc}

 --- F.\ I.\ Santos Cruz's  
[{\it  On the Prostitution at Lisbon}] \cite{Santos1841}.
\vspace{0.35pc}

\noindent {\bf 1854} --- A.\ Oliveira Marreca's  [{\it A Review  and  Memoir on Mr. Alexandre Herculano proposal, requiring  that  the section of economic and administrative  sciences establishes a Project on  Stadistics}] \cite{Marreca1854}. $<FLUL>$
\vspace{0.35pc}

  --- A.\ J. d'\'{A}vila's  [{\it  Report on the Works of the Congress of Statistics held in Brussels  in  1853}] \cite{Avila1854}. \vspace{0.35pc}

\noindent {\bf 1856} ---  A.\ Moreau de Jonnès'  {\it Compendio Estadistico ou Resumo dos Elementos de Estadistica} \cite{Jonnes}, translation of  {\it \'{E}l\'{e}ments de Statistique: comprenant les principes g\'{e}n\'{e}raux de cette science, et un aper\c{c}u historique de ses progr\`{e}s} \cite{Jonnes1847}.
$<ISA>$
\vspace{0.35pc}

\noindent {\bf 1857} --- J.\ Dias Ferreira's  [{\it Essay on the First Elements of  Stadistics   Theory}] \cite{Ferreira1857}.  $<FDUL>$
\vspace{0.35pc}
     
\noindent {\bf 1864} --- creation of the Dire\c{c}\~{a}o-Geral de Estat\'{i}stica [General Direction of Statistics].

--- A.\ J. d'\'{A}vila's  [{\it Report on the Works of the Congress of Statistics held in Berlin  in  1863}], \cite{Avila1864}. 
$<FDUL>$
   \vspace{0.35pc}
   
   --- F.\ P.\  Azevedo's   [{\it  History of prostitution and sanitary police at Porto, followed by a statistical essay on the two elapsed years, comparative tables, etc.}] \cite{Azevedo1864}.
$<FMUL>$
\vspace{0.35pc}
   
\noindent    {\bf 1865} --- The teaching of elementary  Probability Calculus was included in the program of Infinitesimal Calculus,   at the Escola Polit\'{e}cnica de Lisboa [Lisbon Polytechnic School].
   \vspace{0.35pc}

\noindent {\bf 1866} --- D.\ A.\ da Silva's  [Average yearly amortizations at the foremost survival Montepio Mutualist Associations in Portugal]\cite{Silva1866}. 
\vspace{0.35pc}

\noindent {\bf 1870} --- D.\ A.\ da Silva's  [Contributions for the comparative study of the population movements in Portugal]\cite{Silva1870}. $<FDUL>$
\vspace{0.35pc}

\noindent  {\bf 1875} --- G.\ A.\ Pery's    [{\it Geography
and General Statistics of Portugal and the Colonies}]  \cite{Pery1875}.
$<IICT>$   
\vspace{0.35pc}

\noindent {\bf 1887} --- J.\ F.\ Laranjo's [{\it Program for the Chair of Political Economy and Statistics at the Faculty of Law of the University of Coimbra}] \cite{Laranjo1887}. $<FLUL>$
\vspace{0.35pc}

\noindent {\bf 1894} --- M.\ E.\ Garcia's [{\it Notes of Some Magistral Lectures in the Course of Political Economy and Statistics}] \cite{Garcia1894}.
\vspace{0.35pc}

\noindent {\bf 1898} ---  Sid\'{o}nio Paes' [{\it Introduction to the Theory of Observational Errorss}] \cite{Paes1898}. \vspace{0.35pc}

\noindent {\bf 1900} --- Vicomte de  Wildik's [{\it Notice Statistique sur le Portugal}]. \cite{Wildik1900}
$<FLUL>$

   \section{20th Century --- From Rodolpho Guimarães (1904) to 1974}
   
 Karl Pearson, William Sealy Gosset ({\it Student}), Ronald Fisher, Jerzy Neyman, and Egon Pearson were major figures in the early developments of  Mathematical Statistics, namely  in what regards  inference about parameters assuming normality, discriminant analysis, and experimental design. These concepts were gradually imported, especially through lecture notes (some published as books) by Varennes e Mendon\c{c}a at ISA (Instituto Superior de Agronomia), and by Leite Pinto  and by Bento Murteira at ISCEF (Instituto  Superior  de Ci\^{e}ncias Econ\'{o}micas e Financeiras).  On the other hand, the new methods of Statistics were addressed  in a series of  publications on ``methodological issues'' by W.\ L.\ Stevens (a disciple of Fisher who worked in the Anthropology Institute of  Coimbra  between 1942 and 1945), and by his patron  E.\ Tamagnini, and by agronomists influenced by Fisher's school. 
 
 Outside the university context, Guimar\~{a}es \cite{Guimaraes1931} used statistical tests to investigate  public education issues, and Amaro Guerreiro \cite{Guerreiro1947}\footnote{\, An interesting analysis of this book can be found at 
%\href{https://www.deepdyve.com/lp/wiley/manual-de-estat-stica-vwDem3M8IE?utm_source=freeShare&utm_medium=link&utm_campaign=freeShare}{https://www.deepdyve.com/lp/wiley/manual-de-estat-stica-vwDem3M8IE?utm\_source=freeShare\&utm\_medium=link\&utm\_campaign=freeShare}
\url{https://www.deepdyve.com/lp/wiley/manual-de-estat-stica-vwDem3M8IE?utm_source=freeShare&utm_medium=link&utm_campaign=freeShare}.}, a senior civil  servant at INE (Instituto Nacional de Estat\'{i}stica) included in his book a description of Pearson distributions (inserts between pp.\ 246 and 247) and in Chapter 10 (Curve Fitting) in the final section addresses the question of ``Fitting to Pearson Type I Curve"; it   also contains  a chapter (8) on Index Numbers and another (9) on Time Series Analysis.

The   teaching of Probability at the Polytechnic School and later at the Faculty of Sciences of Lisbon   was crucial to making these developments possible.

\subsubsection*{Probability}

Sid\'{o}nio Paes's  1898 thesis \cite{Paes1898}, presented at the Faculty of Mathematics of Coimbra University,   in a way marked the end of the hegemony of the Faculty of Law at the University of Coimbra in what concerns  the university teaching of Statistics in Portugal.

The research, teaching, and dissemination of Probability Calculus effectively enabled the transition from Descriptive Statistics, sufficient for handling official statistics in the 19th century, to the teaching and research of Inferential Statistics.

Since 1865, a chair of Infinitesimal Calculus including the  elementary teaching of Probability Calculus had been taught at the Polytechnic School of Lisbon, and in 1911 a quarterly chair of Probability Calculus was created at the Faculty of Sciences, University of Lisbon, which in 1932 became an annual chair, taught by Victor Hugo de Lemos and Pedro Braumann (cf.\ Fontes de Sousa, 2005 \cite{Fontes}).  The lecture notes \cite{Lemos1945}  from Victor Hugo de Lemos's course was edited in 1945 by J.\ Cabral Madeira.
\bigskip

The five volumes  of Adolpho Loureiro's {\it Os Portos Mar\'{i}timos de Portugal e ilhas Adjacentes: Atlas} \cite{Loureiro1904} [{\it The Sea Ports of Portugal and Adjacent Islands}], containing many statistical tables and details, were published from 1904 to 1909.
\bigskip

\noindent\begin{minipage}{0.22\textwidth}
  \begin{itemize}
 \item []
\includegraphics[width=\linewidth]{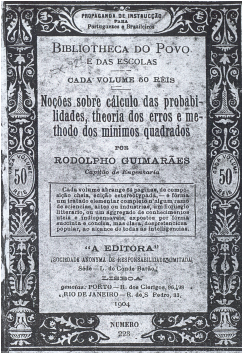}
\end{itemize}
\end{minipage}%
\begin{minipage}{0.7\textwidth}
\begin{itemize}
\item  []  As far as publications in the area of Probability are concerned, the first noteworthy event was the publication of a  booklet intended to the layman, in 1904, volume 223 of the Biblioteca do Povo e das Escolas [Library for the People and the Schools], {\it No\c{c}\~{o}es sobre C\'{a}lculo das Probabilidades, Theoria dos Erros e M\'{e}thodo dos M\'{i}nimos Quadrados}   \cite{Guimaraes1904} [Notions of Probability Calculus,  Theory of Errors  and the Least Squares Method], by Rodolpho Guimar\~{a}es. 
\end{itemize}
\end{minipage}
\bigskip

As Rui Santos   \cite{Santos2008a,Santos2024a,Santos2024b}  comments, Rodolpho Guimar\~{a}es\footnote{~  Rodolfo Guimarães (1866-1918) was a Colonel  in the Portuguese army   and a professor at the Army School of Lisbon (Military Academy). He also served at the Astronomical Observatory of Tapada in Lisbon. Currently, Guimarães is best known for his research in the History of Mathematics, particularly for publishing a catalog of mathematical works by Portuguese authors during the 19th century and for promoting the work of Pedro Nunes (1502--1578), as well as some contributions in the field of Geometry. For detailed information on Rodolfo Guimarães, see Saraiva  \cite{Saraiva}. }, in the scant 64 pages of this small volume, manages to present an interesting introduction to the topic.

The second notable publication is by Jos\'{e} Freire de Sousa Pinto (1855-1911) --- the {\it No\c{c}\~{o}es de C\'{a}lculo das Probabilidades para o Estabelecimento das Bases da Estat\'{i}stica} \cite{SousaPinto1913} [Notions of Probability Calculus for Establishing the Bases of Statistics], analysed in  Santos     \cite{Santos2008b,Santos2016}.  José de Sousa Pinto, a professor at the University of Coimbra, discusses the use of Probability Calculus in modeling random phenomena. In this work, the author begins by explaining the process of constructing knowledge in any experimental science, concluding that the advancement of knowledge results from the combination of observing the phenomenon and theoretical reasoning. Thus, any theory arises from the observation of a phenomenon, initially being an approximation. Then, the phenomenon is observed once more. This new observation serves to test the initial theory, which is confirmed or refined with new knowledge from the new observation, or even altered if the initial theory is found to be erroneous. It is in this successive chain of dual effort, observation and reasoning, that scientific knowledge advances towards a complete understanding of the phenomenon, where the formalized theory will always occur whenever the phenomenon happens.  

It should be noted that the analysis of prerequisites for teaching Statistics was a subject of reflections by Leite Pinto \cite{LeitePinto1945}, in {\it No\c{c}\~{o}es de Matem\'{a}tica necess\'{a}rias ao estudo da Estat\'{i}stica} [Notions of Mathematics Necessary for the Study of Statistics], and by Amaro Guerreiro, in {\it Em Torno do Ensino da Estat\'{i}stica} \cite{Guerreiro1952} [About the Teaching of Statistics], 1952-53.\bigskip

\noindent \begin{minipage}{0.21\textwidth}
  \begin{itemize}
 \item []
\includegraphics[width=\linewidth]{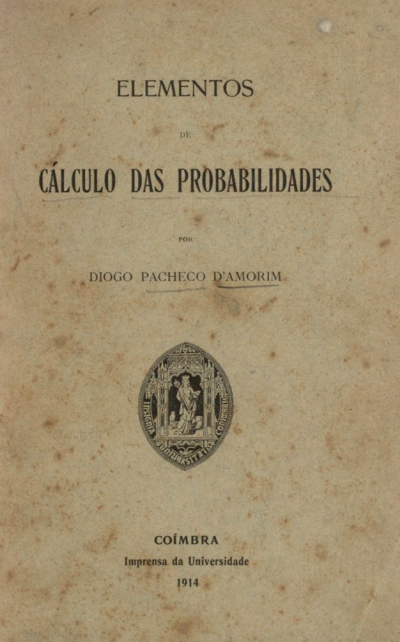}
\end{itemize}
\end{minipage}%
\begin{minipage}{0.6\textwidth}
\begin{itemize}
\item  []  In 1914, Diogo Pacheco d'Amorim (1888-1976) defended his doctoral dissertation at the University of Coimbra on {\it Elementos de C\'{a}lculo das Probabilidades} \cite{PachecoAmorim1914} [Elements of Probability Calculus], the aim of which was to establish rigorous foundations for the theory of Probability. In fact, this objective failed, and another doctoral thesis on Probability \cite{Reis1929}, by Manuel Reis (1929), is very critical of Pacheco d'Amorim's thesis. 
\end{itemize}
\end{minipage}
\begin{minipage}{0.20\textwidth}
  \begin{itemize}
 \item []
\includegraphics[width=\linewidth]{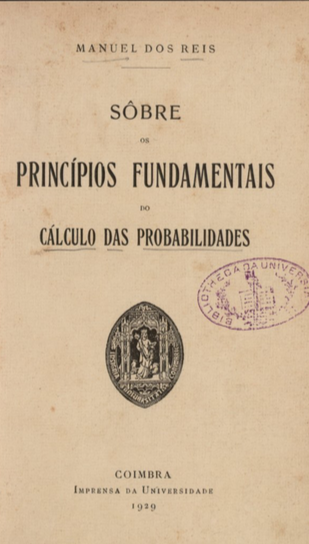}
\end{itemize}
\end{minipage}%
\bigskip

Later, in a unique work by Lu\'{i}s de Albuquerque \cite{Albuquerque1959} on the foundations of the Calculus of Probabilities, those dissertations  are not even mentioned. Pacheco d'Amorim's thesis has been exhaustively studied by Rui Santos    \cite{Santos2008c}, see also \cite{Mendonca}.

See, as a simple introduction to the work of Pacheco d'Amorim, Pestana and Velosa    \cite{PV} and 
Pestana   \cite{Pestana94}. There is an annotated English translation of his doctoral thesis by   Mendon\c{c}a {\it et al.}\  \cite{Mendonca}.  

 In fact, it was only in 1933 that Kolmogorov  \cite{Kolmogorov} made an axiomatic construction of Probability, inspired by Maurice Fr\'{e}chet's ideas on the use of Measure Theory.  \bigskip

\begin{minipage}{0.20\textwidth}
  \begin{itemize}
 \item []
\includegraphics[width=\linewidth]{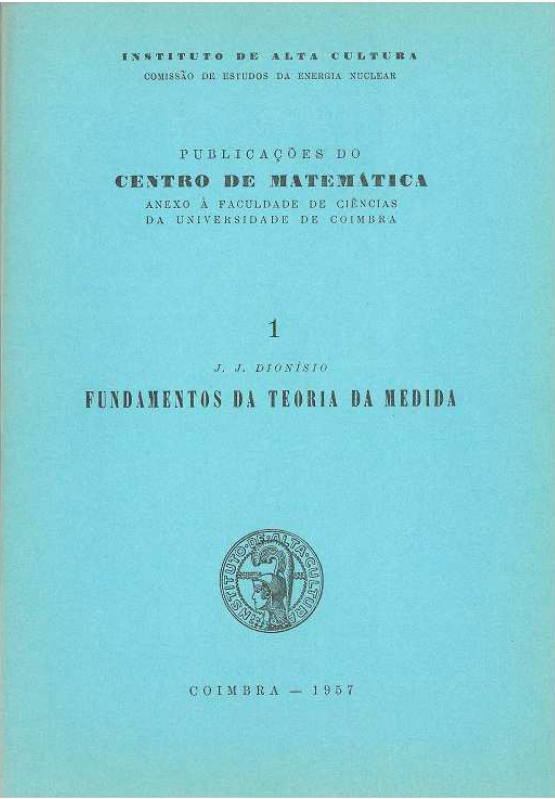}
\end{itemize}
\end{minipage}%
\begin{minipage}{0.5\textwidth}
\begin{itemize}
\item  []  Measure Theory   was tackled by Jos\'{e} Joaquim Dion\'{i}sio \cite{Dionisio1957} (who, in what regards   Probability, published a brief note on Shannon's entropy \cite{Dionisio1959} in {\it Gazeta de Matemática}), and became the preferred field of work  for Pedro Braumann \cite{Braumann1969, Braumann1987}.\end{itemize}
\end{minipage}
\begin{minipage}{0.19\textwidth}
  \begin{itemize}
 \item []
\includegraphics[width=\linewidth]{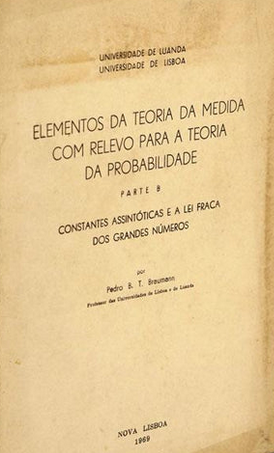}
\end{itemize}
\end{minipage}%
\bigskip

Pedro Braumann, who was a fellow at Stanford in 1955-56, was also the author of an excellent exposition on the central theme of classical Probability Theory, the arithmetic of probability laws, [{\it Introduction to the  Study of Limits of Sums of Independent Casual Variables}] \cite{Braumann1958}.

\subsubsection*{Statistics at the University of Coimbra}
The teaching of Statistics associated  to Political Economy persisted at the Faculty of Law of the University of Coimbra, giving rise to the publication of various texts (Gouveia {\it et al.} \cite{Gouveia1913}, Marnoco e Souza\footnote{\, The incomplete copy from FDUL did not allow us to access some parts of the book; it should be noted that it has a section on ``the sceptics of Statistics'', a subject that is currently widely discussed. In Chapter III (Statistical Laws), sections 59 and 60 are respectively ``Law of large numbers"\ and ``Law of small numbers". Poisson,   wishing that his  result on the convergence of a succession of binomials to Poisson  to be known as the law of small numbers, renamed   ``law of large numbers"   Jacques Bernoulli's ``golden theorem" on the convergence of arithmetic means towards the expectation. Only the expression ``law of large numbers" persisted.} \cite{MarnocoSouza1913}, Silva and Rocha  \cite{SilvaRocha1919}, Rocha and Monteiro  \cite{RochaMonteiro1923}, Pereira  \cite{Pereira1929}).  

On the other hand, at the Anthropology Institute of the University of Coimbra, Eug\'{e}nio Tamagnini had an innovative perspective on the contribution of Statistics in Biometrics research. In {\it Revista da Faculdade de Ci\^{e}ncias da Universidade de Coimbra}, Tamagnini \cite{Tamagnini1937} published an innovative work in Portugal on   analysis of variance. Tamagnini also  hired W.\ L.\ Stevens, a  disciple  of Ronald Fisher. Xavier da Cunha  (1992) \cite{Xavier}    comments  that ``{\it Tamagnini's interest in the statistical treatment of data obtained from various populations and in comparing them led him to hire a foreign professor of statistics, Professor W.\ L.\ Stevens, who collaborated with him on blood groups and also organised free courses in statistics for students of the exact and natural sciences. Stevens had been a brilliant disciple and collaborator of Dr.\ Fisher, having stayed in Coimbra `making us familiar with statistical methods and their most convenient application', in Tamagnini's words. This initiative was undoubtedly very useful.}" 
 
The entry {\it Mathematical Statistics at the University of S\~{a}o Paulo: Mr.\ W.\ L.\ Stevens} \cite{Nature} informs   that ``[W.\ L.\ Stevens] {\it then joined the staff of the Statistical Department, Rothamsted Experimental Station, for a few months on urgent war work prior to leaving for Portugal to take up a lectureship in statistics at Coimbra at the request of the British Council and the Foreign Office. In 1944 he returned to Britain and took a post as statistician with Imperial Chemical Industries, Ltd., Billingham. In 1947 he joined the staff of the Admiralty Statistical Department under Mr.\ H.\ L.\ Seal}", and it can be conjectured that he had something to do with the British intelligence services during the war.  
 
Fonseca  \cite{Fonseca} states that  
``{\it It is in this context of the necessity to improve the statistical treatment of anthropometric, physiological and demographic data, that one understands the arrival of W.\ L.\ Stevens, assigned `as a statistician, to guide the application work of modern statistical methods to Biological Sciences and organise initiation courses for professors and students of the same methods'.  The `Elementary Course of Modern Statistical Methods Applicable to Scientific Investigation', headed by professor W.\ L.\ Stevens, functioned between 1942 and 1944, and is linked to a set of published studies in the series `Quest\~{o}es de M\'{e}todo' [`Methodogical Issues'].
[...] As for the years W.\ L.\ Stevens spent in Coimbra, the data which allows us to inquire about his teaching activity is frankly scarce. We know that the course he led functioned between 1942 and 1944, but the apparent absence of Stevens' curriculum vitae in the archives of the University of Coimbra, and, moreover, the inexistence of annuals of the same institution for the period between 1943 and 1947 (encompassing, thus, the years during which the course led by W.\ L.\ Stevens was lectured), conditioned our investigation. What was the programme of the course? Who attended it? What bibliography was recommended? Why was it extinct after only two years? These are just some of the questions that (at least) for now, will remain without an answer.}".   

Between 1942 and 1945 Stevens published a remarkable set of works which he categorised as Questions of Method --- namely on the $\chi^2$ significance test, discriminant analysis, statistical distributions, parameters estimation ---, disseminating innovations in mathematical statistics by Karl Pearson, `Student', Ronald Fisher, Egon Pearson, Jerszy Neyman and their schools.  

Stevens also published, with Aureliano Quintanilha and Hugo Ribeiro \cite{Quintanilha1942}  an interesting article on the application of Probability Calculus in the investigation of non sexual reproduction by plasmogamy followed by karyogamy in the higher fungi basidiomycetes, in the
{\it Gazeta de Matem\'{a}tica} (both in the index and in the article the initials W and L are interchanged).

\subsubsection*{Statistics at the Technical University of Lisbon}
Several graduates in Mathematical Sciences, Mathematics and Geographical Engineering from FCUL, where they had studied Probability, Errors and Statistics, became professors at institutions that developed the teaching of Mathematical Statistics, namely at Institutes of the Technical University of Lisbon. Special mention should be made of 
\begin{itemize}
\item Manuel Zaluar Nunes (1907-1967), who started as an assistant at ISCEF,  before going to study in Paris as a fellow of the Instituto de Alta Cultura, published in 1933 with M\'{a}rio Santos a concise exposition of {\it Elementos de C\'{a}lculo das Probabilidades e de Estat\'{i}stica Matem\'{a}tica} \cite{SantosZaluar1933}  [{\it Elements of Probability Calculus and Mathematical Statistics}]: written in accordance with the program of the 2nd chair of the Instituto Superior de Ci\^{e}ncias Econ\'{o}micas e Financeiras. Founder of the {\it Gazeta de Matem\'{a}tica}, of {\it Portugaliae Mathematica}, and of the Sociedade Portuguesa de Matem\'{a}tica  [Portuguese Mathematical Society], he was an Assistant Professor at FCUL and later a Full Professor at ISA, a position from which he was dismissed in 1947 for political reasons, having to migrate first to France, where he was a research assistant at the Centre National de la Recherche Scientifique, and later to Brazil, where he was hired as a professor at the University of Recife.

\item Francisco Paulo Pinto Leite (1902-2000)  was   lecturer at   Sorbonne from 1931 to 1933. One of the founders of the Portuguese Mathematical Society, he was appointed Full Professor at ISCEF in 1940, retiring in 1973. There are various versions of the course notes for the Statistics course he taught at ISCEF, and he was also the first lecturer, in 1952-53, of the chair of Econometrics, with Henri Guitton (University of Dijon) and Jos\'{e} de Castañeda (University of Madrid), having as assistant Manuel Jacinto Nunes (further information in  Machado and  Santos Silva   \cite{MachadoSilva}).

It should be noted that 1953 saw the publication of {\it Trabalhos do Semin\'{a}rio de Econometria Dirigido pelo Prof.\ H.\ O.\ Wold} \cite{Wold1953b} [{\it Works presented at the Seminar of Econometry drected by Professor H.\ O.\ Wold}] and the translation of Wold's book on dynamic systems \cite{Wold1953b}  by Jacinto Nunes, and that Mira Fernandes (who prefaced the book by M\'{a}rio Santos and Zaluar Nunes) consistently supported research in Statistics and Econometrics, see the series of eight volumes of {\it Estudos de Matem\'{a}tica, Estat\'{i}stica e Econometria} [{\it Studies in Mathematics, Statistics and Econometrics}] \cite{MiraFernandes1956}  that he edited between 1956 and 1964. 

\item Jos\'{e} Sebasti\~{a}o e Silva (1914-1972), a fellow of the Instituto de Alta Cultura in Rome from 1942 to 1946, in 1951 won the public competition for the position of Full Professor at the Instituto Superior de Agronomia, taking charge of the chairs of General Mathematics and of Calculus of Probabilities. In 1954-55, he prepared didactic texts, collected in the course notes {\it C\'{a}lculo das Probabilidades}\cite{SebastiaoSilva1954}  [{\it Calculus of Probabilities}]  published at ISA, with an appendix from 1957-58 on correlation and regression, adjustment by the method of least squares, and Student's $t$ and Fisher-Snedecor's $F$ statistics, cf.\ Neves    \cite{Neves}, pp.222--223. In 2000  Calouste Gulbenkian Foundation republished this work \cite{SebastiaoSilva1999}, now unfortunately out of print.
\end{itemize}

Perhaps as a result of the influence of the work developed by Sir Ronald Fisher and his collaborators at Rothamsted, new areas of Statistics were developed at ISA, particularly in the area of Experimental Design (Costa \cite{Costa1933}, Carvalho  \cite{Carvalho1945, Carvalho1946}) and Linear Models (Pestana \cite{Pestana1957}).

\begin{itemize}
\item
Professor Varennes e Mendon\c{c}a, who graduated in Agronomy Engineering at ISA,  taught Probabilities at ISA before the appointment of Professor Sebasti\~{a}o e Silva, publishing a brief course on {\it No\c{c}\~{o}es de C\'{a}lculo das Probabilidades} \cite{Varennes1950}  [Notions of Probability Calculus], for details cf.\ Neves  \cite{Neves}, pp.\ 224--225.  Varennes e Mendon\c{c}a early on realized the difficulties of translating statistical terms, publishing an interesting article \cite{Varennes1942} on terminology in 1942.
\end{itemize}
\smallskip

\noindent   \begin{minipage}{0.25\textwidth}
  \begin{itemize}
 \item []
\includegraphics[width=\linewidth]{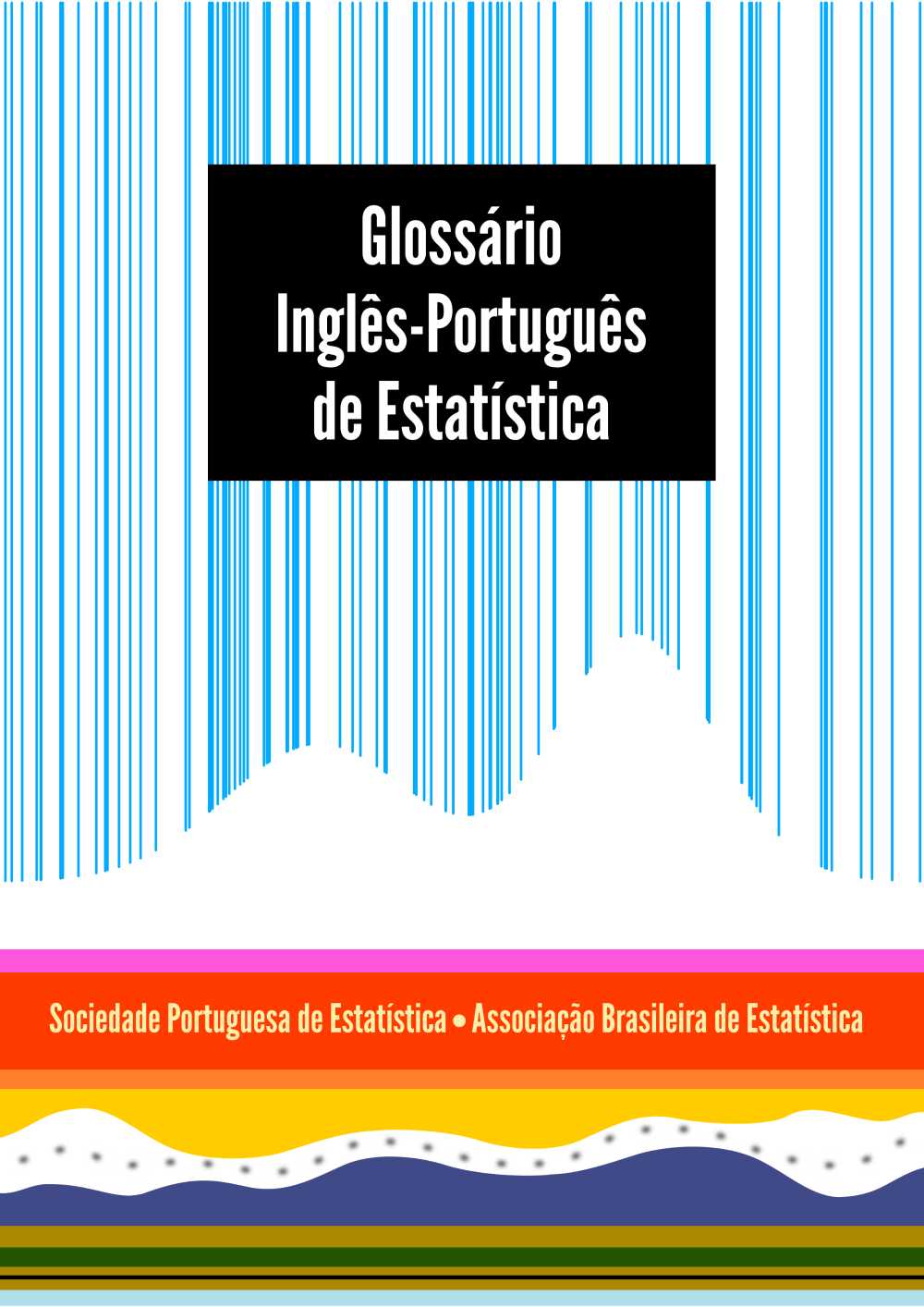}
\end{itemize}
\end{minipage}%
\begin{minipage}{0.7\textwidth}
\begin{itemize}
\item  []  Murteira and Madureira \cite{MurteiraMadureira1958} supplied  interesting translations, but it wasn't until much later that the Specialised Committee on Statistical Nomenclature of the Portuguese Statistical Society and the Brazilian Statistical Association%, whose members were C.\ D.\ Paulino (President), D.\ Pestana, J.\ Branco,L.\ Carvalho; A.\ E.\ dos Reis, J.\ Singer, L.\ Barroso andW.\ Bussab,  
 prepared the 
{\it Statistics Glossary
English-Portuguese} \cite{Paulino2006},  
%Portuguese Statistical Society / Brazilian Statistical Association
%\href{https://spestatistica.pt/glossario}{https://spestatistica.pt/glossario}
\url{https://spestatistica.pt/glossario}, which is periodically updated. See also Ventura \cite{Ventura} {\it et al}'s 
[{\it Glossary of Statistical Terms: German, French, English, Portuguese}]. 
\end{itemize}
\end{minipage}
\smallskip

\subsubsection*{Publishers invest in Statistics --- translations and dissemination}

The translation, in 1945,  of Snedecor's  {\it Statistical Methods Applied to Experiments in Agriculture and Biology}  \cite{Snedecor1945} certainly had a major impact on the development of Statistics in Portugal. The popularisation of interest in Probability and Statistics, with an increase in the number of students taking courses in these areas, finally aroused the interest of publishers.\medskip

\noindent \begin{minipage}{0.24\textwidth}
  \begin{itemize}
 \item []
\includegraphics[width=\linewidth]{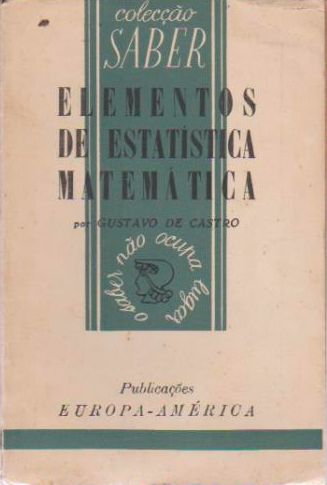}
\end{itemize}
\end{minipage}%
\begin{minipage}{0.5\textwidth}
\begin{itemize}
\item  []   Publica\c{c}\~{o}es Europa-Am\'{e}rica pioneered this move to publish statistical books,  with the publication of Gustavo de Castro's books \cite{Castro1959,Castro1965} in the collection {it Saber} [Knowledge] (1959, 1965), an author chosen probably because he had previously published a book on {\it Infer\^{e}ncia Estat\'{i}stica} \cite{Castro1953} [Statistical Inference] at LNEC [National Laboratory of Civil Engineering] in 1953.
\end{itemize}
\end{minipage}
\begin{minipage}{0.24\textwidth}
  \begin{itemize}
 \item []
\includegraphics[width=\linewidth]{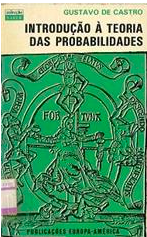}
\end{itemize}
\end{minipage}%
\bigskip

This was followed by the publication of the translation of Tippett's Statistics \cite{Tippett1955}  in 1955 (reprinted in 1968), and in 1973-75 Europa-Am\'{e}rica published the two volumes of P.\ Dagnelie's  {\it Théorie et Méthodes Statistiques: Applications Agronomiques} \cite{Dagnelie1973}, translated by A.\ St.\ Aubyn, and in 1982 the translation of  M.\ L.\ L\'{e}vy's {\it Comprendre les Statistiques} \cite{Levy1982}.\smallskip

\bigskip
\hspace*{-1.5cm}\begin{minipage}{0.2\textwidth}
  \begin{itemize}
 \item []
\includegraphics[width=\linewidth]{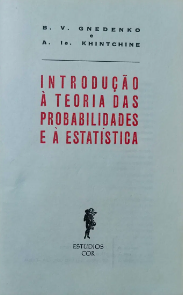}
\end{itemize}
\end{minipage}%
\begin{minipage}{0.21\textwidth}
  \begin{itemize}
 \item []
\includegraphics[width=\linewidth]{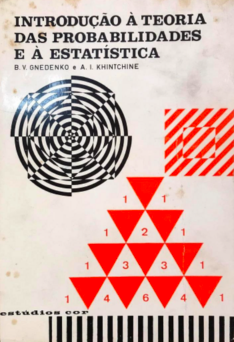}
\end{itemize}
\end{minipage}%
\begin{minipage}{0.43\textwidth}
\begin{itemize}
\item  []  Editorial Est\'{u}dios COR also ventured out with the publication in 1968   of Gnedenko and Khinchine's book  {\it Elementary Introduction to the Theory of Probability} \cite{Gnedenko1968}, with   preface and translation by A.\ Sim\~{o}es Neto.  Despertar published a translation of one of the highest quality and most successful introductory  books, Moroney's {\it Facts from Figures} \cite{Moroney1969}, in 1969. 
\end{itemize}
\end{minipage}
\begin{minipage}{0.2\textwidth}
  \begin{itemize}
 \item []
\includegraphics[width=\linewidth]{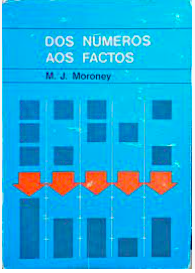}
\end{itemize}
\end{minipage}%

   \subsubsection*{Some highlights}
   
	The books by Guimar\~{a}es \cite{Guimaraes1904} and Sousa Pinto \cite{SousaPinto1913} were analysed by Rui Santos, and those by Varennes e Mendon\c{c}a \cite{Varennes1950}, Sebasti\~{a}o e Silva \cite{SebastiaoSilva1954} and Pestana \cite{Pestana1957} by Manuela Neves \cite{Neves}, as already mentioned. 
   
   Of the others, we think the publications by Santos and Zaluar Nunes \cite{SantosZaluar1933}, Braumann \cite{Braumann1958}, Reis and Sarmento \cite{ReisSarmento1960} and Dias Agudo \cite{DiasAgudo1961} deserve special mention.
   
The publications by E. Tamagnini \cite{Tamagnini1937}  and W.\ L.\ Stevens \cite{Stevens1942,Stevens1944,Stevens1945a, Stevens1945b,Stevens1945c} in the {\it Revista da Faculdade de Ci\^{e}ncias da Universidade de Coimbra} [Review of the Faculty of Sciences of Coimbra University] also deserve special mention.

 \noindent \begin{itemize}
\item The 1933 book by  Santos and Zaluar Nunes \cite{SantosZaluar1933} has a modernity that contrasts sharply with the 1932 book by Bueno y Martins \cite{BuenoMartins1932}, which has a perspective closer to official statistics. \bigskip

\hspace*{-0.8cm} \begin{minipage}{0.21\textwidth}
  \begin{itemize}
 \item []
\includegraphics[width=\linewidth]{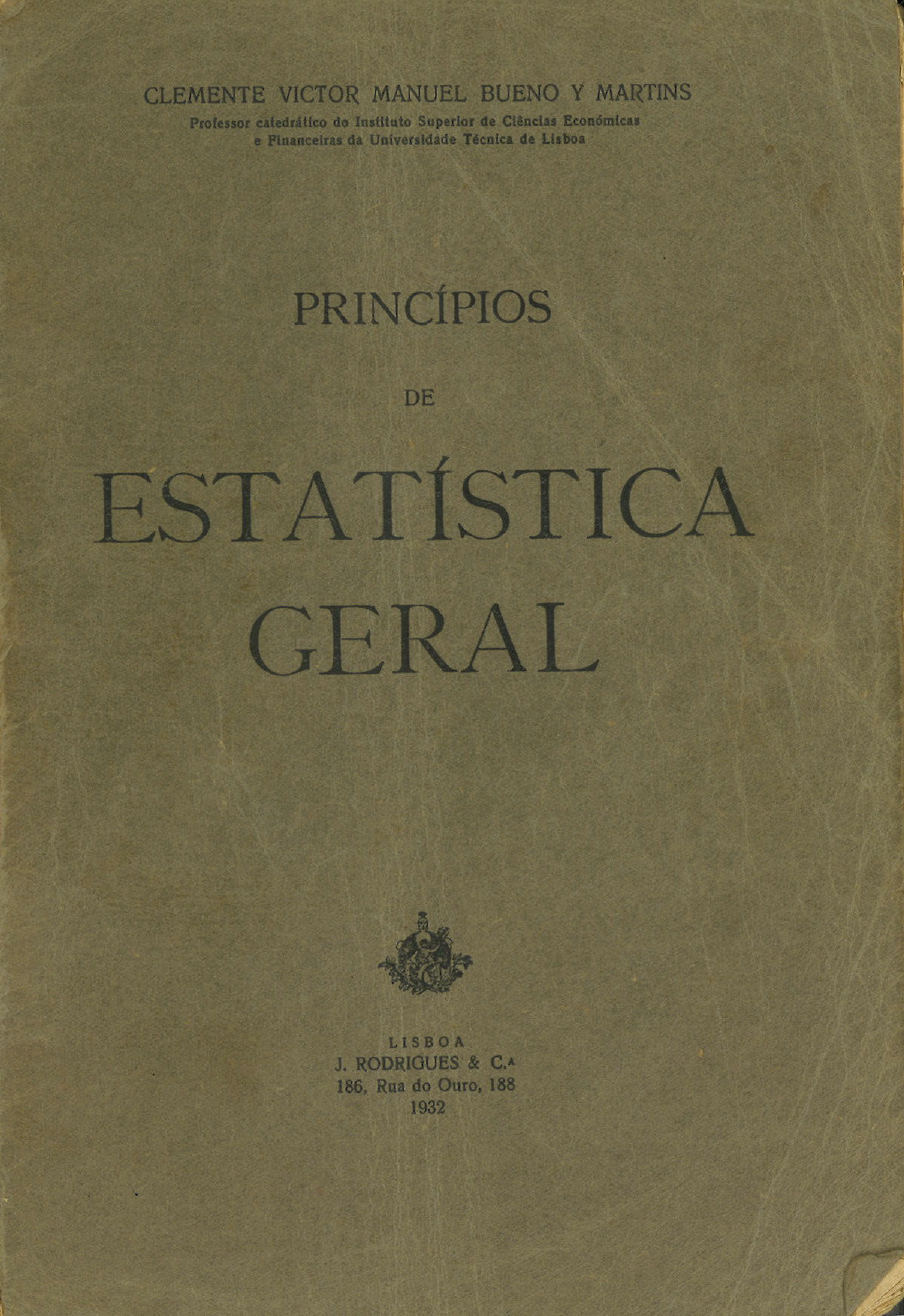}
\end{itemize}
\end{minipage}%
\begin{minipage}{0.5\textwidth}
\begin{itemize}
\item  []  However, when it comes to descriptive statistics, Bueno y Martins' book is more detailed, for example, in addition to the arithmetic mean, he considers geometric, harmonic and weighted averages. 
\end{itemize}
\end{minipage}
\begin{minipage}{0.21\textwidth}
  \begin{itemize}
 \item []
\includegraphics[width=\linewidth]{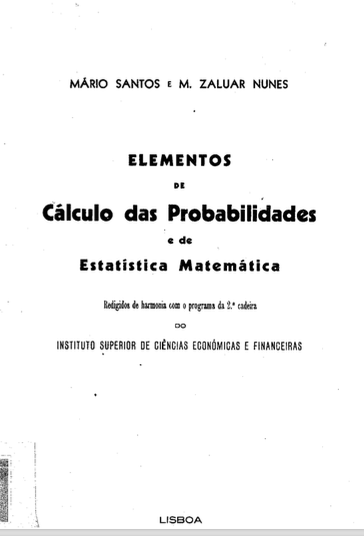}
\end{itemize}
\end{minipage}%
\bigskip

In what regards  graphical representations, Bueno y Martins  discusses cartograms, stereograms, presenting various interesting pictograms, and discusses   as well as association and contingency, and correlation. It should be noted, however, that some of the more elaborate graphics in Bueno y Martins \cite{BuenoMartins1932}  come from other sources. For example, the extra text between pages 138 and 139 is a reproduction of the cover of the August 1930 issue of the {\it Boletim da Dire\c{c}\~{a}o Geral de  Estat\'{i}stica}   [Bulletin of the General Direction of Statistics], see the {\it Cat\'{a}logo Bibliogr\'{a}fico}  [Bibliographic Catalogue], p.19, published by INE in 2013 \cite{Catalog}.  Bueno y Martins' main statistical reference is the monograph by Yule (the bibliographical sources used are indicated at the end of each section, which is very useful), and it is worth noting that Bueno y Martins indicates several books on Probability in the final bibliography, including {\it A Treatise on Probability} by Maynard Keynes.

\bigskip

\noindent \hspace{-1cm}\begin{minipage}{0.63\textwidth}
  \begin{itemize}
 \item [] 
\includegraphics[width=\linewidth, angle=0.91]{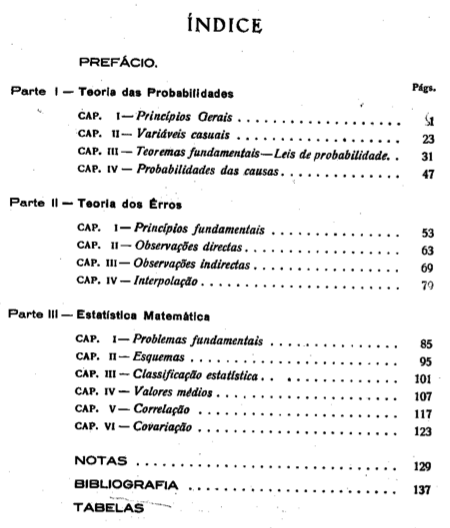}
\end{itemize}
\end{minipage}%
\begin{minipage}{0.35\textwidth}
\begin{itemize}
\item  []\small  {\bf Table of Contents:}\\

{\bf \tiny  FOREWORD.}\\

{\bf  \tiny   Part I --- Theory  of Probability.}\\

\tiny Chapter I --- General Principles.\\
\tiny Chapter II --- Random variables.\\
\tiny Chapter III --- Fundamental theorems --- Probability laws.\\
\tiny Chapter IV --- Probability  of  causes.\\

{\bf \tiny   Part II --- Theory  of Errors.}\\

\tiny Chapter I --- Fundamental  Principles.\\
\tiny Chapter II --- Direct observations.\\
\tiny Chapter III ---  Indirect observations.\\
\tiny Chapter IV --- Interpolation.\\

{\bf \tiny   Part III --- Mathematical Statistics.}\\

\tiny Chapter I --- Fundamental  Problems.\\
\tiny Chapter II --- Schemes.\\
\tiny Chapter III --- Statistical classification.\\
\tiny Chapter IV --- Mean values.\\
\tiny Chapter V --- Correlation.\\
\tiny Chapter VI --- Covariation.\\

{\bf \tiny   NOTES.} \\
{\bf \tiny   REFERENCES.} \\
{\bf \tiny   TABLES.}
\end{itemize}
\end{minipage}

\medskip 
The themes addressed by Santos and Zaluar Nunes are still very elementary, corresponding to a style that has not yet incorporated  the ideas of Pearson, Student and Fisher.   This book was published before Zaluar Nunes, at the time 26 years old, had  his initial period as a $estrangeirado$ in Paris.
 In the bibliographical references, the Probability manuals predominate (Poincar\'{e}, 1912; Borel, 1925;  L\'{e}vy, 1925; Castelnuovo, 1928), and the Statistics references are  Yule (1919), Jordan (1927) and Darmois (1928).

 As far as statistics is concerned, after some elements of descriptive statistics, they discuss what they call Poisson and Lexis schemes, leading to what they consider to be sub- and super-normal, averages, correlation and covariation.
 It's an interesting book,   outstanding among portuguese books on  Statistics   at the time of its publication, but it falls far short of what Leite Pinto, Varennes e Mendon\c{c}a, and Murteira a decade later imported from the new concepts of Mathematical Statistics.

 \item The publications by W.\ L.\ Stevens in {\it Revista da Faculdade de Ci\^{e}ncias da Universidade de Coimbra} [Review of the Faculty of Sciences of the University of Coimbra] are the ones that effectively introduced the new concepts and methods  of Mathematical Statistics in Portugal.
 
 ``Teoria matem\'{a}tica dalgumas distribui\c{c}\~{o}es usadas em estat\'{i}stica'' \cite{Stevens1942} [``Mathematical theory of some distributions used in statistics"], from 1942, discusses random pairs, joint distribution of mean and variance estimators, linear combinations of normal variables, with emphasis on contrasts, and partitioning the $\chi^2$. It introduces Student's $t$ variable, and applies Student's $t$ test to regression.
 
 ``Estima\c{c}\~{a}o estat\'{i}stica'' \cite{Stevens1944} [``Statistical estimation"], published in 1944, is the most interesting from a theoretical point of view, as it introduces the concepts of likelihood, consistency, efficiency, sufficiency, minimum variance and information, always with mathematical details and examples. It also discusses homogeneity and adjustment tests. In the ``Summary and bibliography''\ it refers to the groundbreaking work by Fisher  (1922),  and claims to supply ``{\it for the first time, a demonstration of the methods and formulae}''\ for generalising to the estimation of two or more parameters. He also makes it clear that he ``{\it rejects the inverse probability method and the so-called `Bayes postulate'}.''

 The 1945 publications in vol.\ XIII are of a more applied nature. In ``Aplica\c{c}\~{a}o do teste $X^2$ \`{a} An\'{a}lise da Vari\^{a}ncia'' \cite{Stevens1945a} [``Application of the $\chi^2$ test to the analysis of variance"] Stevens defines the $\chi^2$ variable and the corresponding number of degrees of freedom, establishes the additive property and then deals with the decomposition of the $\chi^2$ and its application to the analysis of variance (a subject already dealt with by Tamagnini in 1937, \cite{Tamagnini1937}). The mathematical treatment is detailed.

 ``An\'{a}lise Discriminante'' \cite{Stevens1945b}  [``Discriminant Analysis"] also has an in-depth mathematical treatment, particularly with regard to the joint distribution of estimators. It discusses the binormal distribution, with remarkable graphs, for example for concentration ellipses. Analysis of variance is discussed in detail.  
  There is a final bibliography, referring to work by Ronald Fisher and the author, but at the begining Stevens  lists Ronald Fisher, H. Hotelling, Bose, Roy, Mahalanobis and Hsu as mentors of the ``new statistical school''.
	
   ``Novos M\'{e}todos para o Estudo da Gen\'{e}tica Humana" \cite{Stevens1945c} [``New Methods for the Study of Human Genetics"] deals extensively with the estimation of proportions, discussing {\it score} tests.  
Curiously, reflecting trends of the time, and in particular those of his mentor Ronald Fisher and his director in Coimbra, E.\ Tamagnini, the bibliography, citing Finney, Ronald Fisher, Haldane and Oldrycht, is all from publications in the {\it Annals of Eugenics}.

\item Pedro Braumann was Victor Hugo de Lemos's assistant in {\it Probabilidades, Erros e Estat\'{i}stica}  [Probability, Errors and Statistics], and during his time abroad at Stanford he had the opportunity to refine his knowledge of Measure Theory and the arithmetic of probability laws, in the specific sense of limit laws of sums of random variables, and the inverse problem of decomposing a random variable into independent summands. 

His 1958 book, {\it Introdu\c{c}\~{a}o ao Estudo dos Limites de Somas de Vari\'{a}veis Casuais Independentes} \cite{Braumann1958} [An Introduction to the Study of Limits of Sums of Independent Casual Variables], deals with stable, self-decomposable and infinitely divisible laws and canonical representations, which were the great ``inventions'' of Paul L\'{e}vy, Kolmogorov and Khinchine, in a rigorous and complete way.

\item The {\it Manual de Estat\'{i}stica M\'{e}dica} \cite{ReisSarmento1960} [Medical Statistics Handbook] by Carlos Santos Reis and Alexandre Sarmento (1960) is most probably the work of self-taught users of Statistics  who have studied hard. There are 54 items in the bibliography, some of which (Arkin and Colton, 1956; Bancroft, 1957; Croxton, 1953; Darmois, 1952; Mather, 1959; Snedecor, 1959; Weatherburn, 1947) are well-known statistics texts. It covers an extensive list of topics, 
see the table of contents  reproduced in Appendix 2, from the perspective of the application of Statistics in biomedical areas, and with less emphasis on the concepts of the mathematical theory of Statistics. It is certainly out of date for the actual  basic training in Statistics in the area of Life Sciences, but at the time its publication it was top of the range. 

\item When Dias Agudo returned from his time abroad in the USA, the Biology students in FCUL asked him to teach them a course in Statistics, covering  the most necessary results for biologists.\bigskip

\hspace*{-0.8cm} \begin{minipage}{0.69\textwidth}
\begin{itemize}
\item  [] The notes for {\it Introdu\c{c}\~{a}o \`{a} Estat\'{i}stica} \cite{DiasAgudo1961} [An Introduction to Statistics], which run to just over 100 pages, discuss probability, random variables, Binomial and Poisson, Normal, the approximation of the central limit theorem; with regard to Statistics itself, there is a brief chapter on Descriptive Statistics, another on estimating the mean value and variance, and finally a chapter on hypothesis testing (mean value, goodness of fit, and comparing two populations),  see the table of contents in Appendix 3.  Concise and clear, like the masterclasses  Professor Dias Agudo taught. Unfortunately, the students who edited  the notes (Moreira Campos, Waldemar Nunes) didn't include a bibliography.
\end{itemize}
\end{minipage}
\begin{minipage}{0.28\textwidth}
  \begin{itemize}
 \item []
\includegraphics[width=\linewidth]{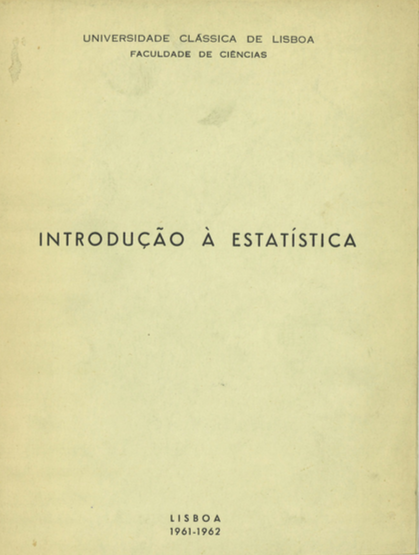}
\end{itemize}
\end{minipage}%

\end{itemize}
 \bigskip 
   
\subsubsection*{Foreigners ...}

Pedro Teodoro Braumann, born in Munich in 1919, left dangerous Nazi-dominated Germany in 1935 and became a naturalised Portuguese citizen. He graduated (1943) and obtained his doctorate (1951) in Mathematics at FCUL, where he was later a Full Professor,   distinguishing himself in teaching and research in Probability Theory. He also taught at the University of Luanda and the University of Aveiro, and ended his career at the FCT of the New University of Lisbon, where he retired in 1989.

As already mentioned, W.\ L.\ Stevens was hired by the University of Coimbra between 1942 and 1944, publishing in Portuguese, in {\it Revista da Faculdade de Ci\^{e}ncias da Universidade de Coimbra} [Review of the Faculty of Sciences of the University of Coimbra] a remarkable set of papers on ``Quest\~{o}es de M\'{e}todo'' [Methodological Issues], which are among the first works on Mathematical Statistics published in Portugal. He also taught free courses in Statistics for students of Exact and Natural Sciences. Unfortunately, the innovative statistical ideas in his work had little influence outside the field of Physical Anthropology.

As it has already been mentioned, in 1953 Henri Guitton (University of Dijon) and Jos\'{e} de Casta\~{n}eda (University of de Madrid) collaborated in the inaugural course in Econometrics at ISCEF, with Leite Pinto and Jacinto Nunes.

\subsubsection*{... and the \textit{estrangeirados} circa 1950s}

Gustavo de Castro, who graduated in Mathematics from FCUL in 1942, was a scholarship holder in Paris from 1946 to 1948, where he completed the course at the Institut de Statistique, obtaining the highest grade in 1948. Upon returning to Portugal, he worked as a researcher at LNEC and ended his professional career as a professor of Biomathematics at the Faculty of Medicine of Lisbon. In the prestigious collection {\it Saber} [Knowledge] of Publications Europa-Am\'{e}rica, he published two popular works on Mathematical Statistics \cite{Castro1959} and on Probability \cite{Castro1965}, as previously mentioned.

In 1953, Bento Murteira (1924-2018) passed his doctoral exams on autoregressive processes, supervised by Maurice Kendall, the celebrated co-author of the bible {\it Advanced Theory of Statistics}. In London, he also interacted with Tippett and Quenouille, the inventor of the jackknife resampling method. After Leite Pinto, he renewed the teaching of Probability, Statistics, and Econometrics at ISCEF, and his lecture notes widely disseminated Mathematical Statistics in Portugal. Many of his publications were collected in the volume {\it Jubileu -- Bento Murteira} \cite{Murteira1994}  published by ISEG in 1994.

In the mid-1950s, several Portuguese mathematicians spent a year abroad in the USA. Regarding Statistics, Tiago de Oliveira's internship with one of the leading experts in extreme values, Emil Julius Gumbel, at Columbia University, substantially changed his scientific life, which initially focused on Algebra. His Aggregation exam was based on substantial work on Estat\'{i}stica de Densidades; Resultados Assint\'{o}ticos \cite{TiagoOliveira1963} [Statistical Distributions; Asymptotic Results], marking  his shift towards teaching and research in Probability and Statistics. His 1956 lecture notes on these subjects were the genesis of {\it Probabilidades e Estat\'{i}stica --- Conceitos Fundamentais} \cite{TiagoOliveira1967} [Probability and Statistics --- Fundamental Concepts], published in 1967-1969 by Livraria Escolar Editora, later republished in 1990-91 by McGraw-Hill with the title {\it Probabilidades e Estat\'{i}stica: Conceitos, M\'{e}todos e Aplica\c{c}\~{o}es} [Probability and Statistics: Concepts, Methods and Applications]. He became one of the most cited Portuguese mathematicians, particularly due to his work in the area of Extreme Value Theory.

For more details, see the {\it Boletim Informativo da Sociedade Portuguesa de Estat\'{i}stica} [Bulletin of the Portuguese Statistical Society], Special Issue in Honor of Professor Jos\'{e} Tiago de Oliveira, December 22, 1998 \cite{Bol2}; also, Gomes  \cite{Gomes93}, The scientific work of J.\ Tiago de Oliveira, and  Gomes  \cite{Gomes2023}, The School of Extremes in Portugal --- PORTSEA, from English ``Portuguese School of Extremes and Applications". 
%\href{https://comum.rcaap.pt/bitstream/10400.26/45255/3/Gomes_2023_A_Escola_de_Extremos.pdf}{https://comum.rcaap.pt/bitstream/10400.26/45255/3/Gomes\_2023\_A\_Escola\_de\_Extremos.pdf}

Pedro Braumann's period at Stanford was also significant for the development of teaching and research in Probability in Portugal. Braumann's monograph \cite{Braumann1958} is a perfect treatise on the classic issues of Probability Theory, and his recurring work on Measure Theory  \cite{Braumann1969, Braumann1987} excellently expounds the fundamentals of Probability Theory.

As previously mentioned, F.\ R,\  Dias Agudo, whose teaching and research career was more related to Linear Algebra, Mathematical Analysis, and Differential Equations, taught an interesting introductory Statistics  course \cite{DiasAgudo1961} at FCUL upon his return from the USA.

\subsubsection*{Chronology}   
    (In regards to lecture notes, we only indicate one of the years of publication)
    
\noindent    {\bf 1904} --- Guimarães' [{\it Notions about Probability Calculus, the Theory of Errors and the Least Squares Method}], \cite{Guimaraes1904}.    
\vspace{0.35pc}

\noindent    {\bf 1904--1909} --- Loureiro's  [{\it The Sea Ports of Portugal and Adjacent Islands}], \cite{Loureiro1904}.

\noindent  {\bf 1909} --- Duncker's  {\it O Méthodo Estatistico da Variação}, translation of   [{\it Methode der Variations-Slatistik}], \cite{Duncker1909}.
\vspace{0.35pc}

\noindent    {\bf 1910} ---  Gracias' [{\it A Primer on Statistics}], \cite{Gracias1910} $<IICT>$
\vspace{0.35pc}

\noindent    {\bf 1911} ---  A quarterly chair of Probability Calculus was created at the Faculty of Sciences, University of Lisbon.

\noindent    {\bf 1913} ---  Sousa Pinto  [{\it Notions of Probability Calculus for Establishing the Bases of Statistics}], \cite{SousaPinto1913}
$<IST>$ 
\vspace{0.35pc}

  --- Gouveia {\it et al.}'s [{\it Lessons of Statistics}],  \cite{Gouveia1913} $<FDUL>$
\vspace{0.35pc}

  --- Marnoco e Souza's  [{\it A Course of Statistics}] (lecture notes compiled by    José Fortes Martinho Simões) \cite{MarnocoSouza1913}.
 $<FDUL>$, incomplete. 
\vspace{0.35pc}

\noindent    {\bf 1914} ---  Pacheco d'Amorim's  [{\it Elements of Probability Calculus}], \cite{PachecoAmorim1914,Mendonca}
\vspace{0.35pc}

\noindent    {\bf 1919} ---  Silva and  Rocha's
 [{\it Statistics; summary of the lectures and questions concening practical problems}]  1913--1919, \cite{SilvaRocha1919}, Manuscript $<FDUL>$ 
\vspace{0.35pc}

\noindent    {\bf 1923} ---  Rocha and  Monteiro's   [{\it Statistics}], 1922--1923, \cite{RochaMonteiro1923},  Manuscript $<FDUL>$ 
\vspace{0.35pc}

\noindent    {\bf 1929} ---  Reis'   [{\it Probability Calculus}], \cite{Reis1929}. 

 \vspace{0.35pc}
 
  ---  Pereira's [{\it Principles of Theoretical and Applied Statistics}], \cite{Pereira1929}. $<ISEG/UL>$
\vspace{0.35pc}

\noindent    {\bf 1931} --- Guimarães' [{\it Tests: Elementary notions of statistical  calculus with applications in scholar measurements}],  \cite{Guimaraes1931}. $<FLUL>$
 \vspace{0.35pc}
 
\noindent    {\bf 1932} --- An annual  chair of Probability Calculus was created at the Faculty of Sciences, University of Lisbon

--- Bueno y Martins' 
 [{\it Principles of General Statistics}] \cite{BuenoMartins1932}.
\vspace{0.35pc}

\noindent    {\bf 1933} ---  Santos  and  Zaluar Nunes'   
 [{\it Elements of Probability Calculus and Mathematical Statistics}], \cite{SantosZaluar1933}. $<ISEG/UL>$
 \vspace{0.35pc}
 
  ---   Costa's   [{\it The Technique of Statistical  Interpretation  of Field Experiments and Mitscherlich Law}], \cite{Costa1933}.
$<ISA>$
 \vspace{0.35pc}
 
 \noindent    {\bf 1935} --- Vasconcelos' [{\it Elements of Infinitesimal Calculus and Probability Calculus}], \cite{Vasconcelos1935} $<ISA>$
\vspace{0.35pc}

\noindent    {\bf 1937} ---    Tamagnini's 
 [{\it The heterogeneity of variation: analysis of variance}] \cite{Tamagnini1937}.   $<FPCEUL>$
\vspace{0.35pc}

  ---     Brito's    [{\it Statistical Treatment of Geophysical Observations}], \cite{Brito1937} $<FCUL>$
 \vspace{0.35pc}
 
 \noindent    {\bf 1942} ---  Varennes e Mendonça's
  [{\it On the {P}ortuguese statistical terminilogy}], \cite{Varennes1942}. $<ISA>$
\vspace{0.35pc}

  --- Quintanilha {\it et al.}'s   [{\it Application  of  Probability Calculus in solving a problem in Biology}], \cite{Quintanilha1942}.
\vspace{0.35pc}

  --- Stevens' [{\it Mathematical theory of some distributions used in statistics}],  \cite{Stevens1942}.  $<FPCEUL>$
\vspace{0.35pc}

\noindent    {\bf 1944} --- Stevens' [{\it Statistical estimation; the theory of estimation of two or more parameters, exemplified with the problem of estimation of the frequencies of genes of blood groups}], \cite{Stevens1944} $<FPCEUL>$ 
\vspace{0.35pc}

\noindent    {\bf 1945} --- Stevens' [{\it Application of the $X^2$  test to the analysis of variance}], \cite{Stevens1945a} 
(Note: in the Table of Contents  there is an error: vectorial analysis, instead of variance analysis) $<FPCEUL>$
\vspace{0.35pc}

  --- Stevens' [{\it Discriminant Analysis}], \cite{Stevens1945b}.
(Note: in the Table of Contents it is attributed to  Eugénio Tamagnini; but it is clearly  W.\ L.\ Stevens' work.) $<FPCEUL>$\vspace{0.35pc}

  ---Stevens' [{\it New Methods for the Study of Human Genetics}], \cite{Stevens1945c}.
 $<FPCEUL>$
\vspace{0.35pc}

  --- Carvalho's [{\it Statistics; Modern  Methods Appliable to Field Experiments}], \cite{Carvalho1945}  $<ISA>$
\vspace{0.35pc}

  --- Lemos'  [{\it Notes on Probability Calculus}], \cite{Lemos1945} $<FCUL>$
\vspace{0.35pc}

\noindent  {\bf 1945} ---     Snedecor's  {\it Métodos Estatísticos. Aplicados a Experimentação Agrícola e Biológica}, translation of   [{\it Statistical Methods Applied to Experiments in Agriculture and Biology}],  \cite{Snedecor1945}  $<IICT>$
\vspace{0.35pc}

  --- Leite Pinto's    [{\it Notions of Mathematics Necessary for the Study of Statistics}],  \cite{LeitePinto1945}. $<FCUL>$
\vspace{0.35pc}

\noindent    {\bf 1946} ---          Carvalho's  [{\it Statistics  and Agricultural Experimentation}], \cite{Carvalho1946} $<IICT>$
\vspace{0.35pc}

\noindent    {\bf 1947} --- Guerreiro's    [{\it Statistics Handbook}], \cite{Guerreiro1947} . \vspace{0.35pc}

\noindent    {\bf 1949} --- Leite Pinto's lecture notes     [{\it Statistical Lessons}], \cite{LeitePinto1949}.
\vspace{0.35pc}

\noindent    {\bf 1950} --- Varennes e Mendonça's [{\it Notions of Probability Calculus}], \cite{Varennes1950} $<ISA>$
\vspace{0.35pc}

  --- Baptista's 
 [{\it Discriminant Analysis. A Preliminary study with Amygdalus Communis, L.}],   \cite{Baptista1950}
$<ISA>$ 
 \vspace{0.35pc}

\noindent    {\bf 1952} ---  Guerreiro's  [{\it About the Teaching of Statistics}], \cite{Guerreiro1952} $<ISCSP>$
 \vspace{0.35pc}
 
\noindent    {\bf 1953} --- A chair of Econometry was created at ISCEF. \vspace{0.35pc}

---  Madeira's
 [{\it Statistical  Quality Control}], \cite{Madeira1953}.
 \vspace{0.35pc}
 
  ---  Castro's [{\it Statistical Inference}], \cite{Castro1953}. \vspace{0.35pc}

  --- Soares da Veiga and  Ponte's   [{\it Introduction to the study of economic statistics}], \cite{SoaresPonte1953}.\vspace{0.35pc}
	
	 --- Murteira's [{\it Some properties of auto-regressive processes}], \cite{Murteira1953}.
\vspace{0.35pc}

 --- Wold's {\it Sistemas Dinâmicos de Tipo Recursivo: Aspectos Económicos e Estatísticos}, translation of  [Recursive Dynamical Systems; Economical and Statistical Aspects],  \cite{Wold1953a}
\vspace{0.35pc}

 --- Wold's  {\it Trabalhos do Seminário de Econometria Dirigido pelo Prof.\  H.\ O.\ Wold} [Works of the Seminar of Econometry directed by Prof.\ H.\ O.\ Wold]  \cite{Wold1953b}.
\vspace{0.35pc}

\noindent    {\bf 1954} ---   Sebastião e Silva's  [{\it Probability Calculus}], \cite{SebastiaoSilva1954} \vspace{0.35pc}

--- Murteira's lecture notes  [{\it Statistics}], \cite{Murteira1954}. \vspace{0.35pc}

\noindent    {\bf 1955} --- Tippett's  {\it Estatística}, translation of  [Statistics], \cite{Tippett1955}, reprinted in 1968.
\vspace{0.35pc}
   
 \noindent    {\bf 1956}  --- Tiago de Oliveira's lecture notes  [{\it Probability, Errors and Statistics}],  \cite{TiagoOliveira1956}. \vspace{0.35pc}
   
--- Pacheco d'Amorim's lecture notes 
[{\it Symbolic Calculus and Calculus of Finite Differences. Probability Calculus}], \cite{PachecoAmorim1956}.\vspace{0.35pc}

 \noindent    {\bf 1956--1964} ---  Mira Fernandes' [{\it Studies in Mathematics, Statistics and Econometry}], 8 volumes, \cite{MiraFernandes1956}.  $<IST>$
  \vspace{0.35pc}
   
\noindent    {\bf 1957} ---   Pestana's [{\it Notes on the Estimation and Significance of the Parameters of Normal Linear Models}].  \cite{Pestana1957}  $<ISA>$
   \vspace{0.35pc}
   
  --- Martinez's  [{\it Introduction to an  Essay on Economical Statistics}], \cite{Martinez1957}  \vspace{0.35pc}
   
  --- Dionísio's  [{\it Foundations of Measure Theory}], \cite{Dionisio1957}.
\vspace{0.35pc}

\noindent    {\bf 1958} ---  Braumann's [{\it An Introduction to the Study of Limits of Sums of Independent Casual Variables}], \cite{Braumann1958}
\vspace{0.35pc}

  ---   Murteira and  Madureira's  [{\it Portuguese Statitical  Terminilogy}],   \cite{MurteiraMadureira1958} $<ISEG>$
   \vspace{0.35pc}
   
  --- Castro's [{\it Elements of Mathematical Statistics}] 
(2nd  edition in  1983, entitled    [{\it Classical Mathematical Statistics ---  the Ideas}]),  \cite{Castro1959}
\vspace{0.35pc}

\noindent  {\bf 1959} --- Albuquerque's  [{\it Notes on the Foundations of Probability Calculus}], \cite{Albuquerque1959}.
\vspace{0.35pc}

  --- Dionísio's  [{\it The definition of entropy in probability calculus}], \cite{Dionisio1959}
\vspace{0.35pc}

 --- Sá's lecture notes 
[{\it Probability, Errors and Statistics}], \cite{Sa1959}. $<FCUL>$
\vspace{0.35pc}

\noindent  {\bf 1960} --- Reis and  Sarmento's  [{\it Medical Statistics Handbook}], \cite{ReisSarmento1960} $<IICT>$
\vspace{0.35pc}

\noindent   {\bf 1961} ---      Dias Agudo's    [{\it An Introduction to Statistics}], \cite{DiasAgudo1961}.\vspace{0.35pc}
  
\noindent   {\bf 1962} ---  Lopes'   lecture notes [{\it Probability, Errors and Statistics}], \cite{Lopes1962}.
\vspace{0.35pc}

\noindent  {\bf 1963} --- Tiago de Oliveira's [{\it Statistical Distributions; Asymptotic Results}], \cite{TiagoOliveira1963}.\vspace{0.35pc}

\noindent  {\bf 1965} --- Castro's  [{\it Introduction to the Theory of Probability}], \cite{Castro1965}. 
\vspace{0.35pc}

\noindent  {\bf 1967} --- Tiago de Oliveira's [{\it Probability and Statistics  -- Fundamental Concepts}], \cite{TiagoOliveira1967}.
\vspace{0.35pc}

\noindent  {\bf 1968}
  --- Gnedenko and Khintchine's {\it Introdução à Teoria das Probabilidades e à Estatística}, translation of  [An Elementary Introduction to the Theory of Probability],  
\cite{Gnedenko1968}.
\vspace{0.35pc}
 
 \noindent  {\bf 1969} ---  Braumann's [{\it Elements of Measure Theory Relevant for Probability Calculus}], \cite{Braumann1969}.\vspace{0.35pc}

 --- Moroney's   {\it Dos Números aos Factos}, translation of  [Facts from Figures],   \cite{Moroney1969}.
 \vspace{0.35pc}

\noindent  {\bf 1971--1973} --- Galvão de Mello's  [{\it Introduction to Statistical  Methods}], \cite{GalvaoMello1971}. (2nd ed.  1993-1997 entitled  [{\it Probability and Statistics: Concepts and Fundamental Methods}].

   \section{The last 50 years}
   
 Fifty years is too short a time to have a historical perspective, and in this section, we will only highlight a few issues that seem to us to be more relevant.

Veiga Simão, who distinguished himself as Rector of the General University Studies of Mozambique between 1962 and 1970, was Minister of Education from 1970 to 1974. He profoundly altered higher education in Portugal, and many of the changes following April 25, 1974, are still a consequence of the deep reform he initiated.

In 1973, the Nova University of Lisbon, the University of Aveiro, the University of Minho, the University Institute of \'{E}vora, and the Polytechnic Institute of Covilhã were established. In 1979, the last two  became the University of \'{E}vora and the University of Beira Interior. The University of the Azores was created in 1976, the University of Trás-os-Montes and Alto Douro in 1986, and the Open University and the University of Madeira in 1988. Starting in 1979, fifteen public Polytechnic Institutes were established. The 1986 Basic Law of the Education System clarified the objectives of university and polytechnic higher education and the degrees and titles they confer, necessitating the hiring of qualified teaching/research staff.

The political changes of 1974 initially caused a deep crisis in universities, but it is worth remembering that crisis and opportunity form an interesting binary. There was a strong investment in hiring teaching staff and sending young researchers abroad for obtaining their doctorates. By 2000, 38 university teachers from Portugal had obtained doctorates in Statistics abroad, and 4 obtained doctorates in Portuguese universities supervised by foreigners (with an additional 4 doctorates supervised in Portugal, but by assistants who had previously spent a period of postgraduate studies or internships abroad), cf.\ Amaral Turkman  \cite{Anton}.

The Portuguese Statute of the University Teaching Career (Decree-Law 448/79, November 13, ratified by Law 19/80 of July 16) doubled the university staff (expanded with a board of supernumerary staff when necessary, later added to the regular staff). Decree-Law 66/80, of April 9, established the rules for creating departments and autonomous sections in universities, and Decree-Law 263/80, of August 7, legislated the creation of Master's degrees.
   
Due to the dynamism and prestige of J. Tiago de Oliveira, the Center for Statistics and Applications of the University of Lisbon (CEAUL) was created in 1975, joined by researchers from other universities, and in the Bachelor's degree in Applied Mathematics, courses in Stochastic Processes, Computational Statistics  and Simulation were created.

In the early 1980s, FCUL had the critical mass\footnote{\, Doctorates in Sheffield, M.\  Ivette Gomes (1978), Dinis Pestana (1978), Antónia Amaral Turkman (1980), and Feridun Turkman (1980); and also Cristina Sernadas (London, 1980) and Helena Nicolau and Fernando Nicolau (1981), supervised by Tiago de Oliveira after obtaining the \textit{doctorat de troisième cycle} in Paris.} to create the Department of Statistics, Operational Research, and Computing in 1981 (with Computing soon separating), and a Master's course in Statistics and Operational Research. New courses were also created, such as Order Statistics, Computational Statistics, Sampling, Quality Control, Design of Experiments, Reliability, and Non-Parametric Methods.

  Starting in the mid-1970s, the University of Coimbra substantially developed the teaching of Statistics with the collaboration of foreign professors, namely Quidel, Delecroix, Moché, Deheuvels, and Teugels. The doctorates of Nazaré Mendes Lopes (1985, supervised by Geffroy), Esmeralda Gonçalves (Lille, 1988), Paulo Oliveira (Lille, 1991), Ana Cristina Rosa (Toulouse, 1994), Emília Nogueira (1993, supervised by Nazaré Mendes Lopes and Delecroix), and Carlos Tenreiro (1995, supervised by Nazaré Mendes Lopes and Gouriéroux) strengthened the area of Statistics in the Department of Mathematics, which housed the Center for Mathematics of the University of Coimbra,  homologated by INIC in 1978, with a research line in Probability, Statistics, and Stochastic Processes, directed by Maurice Moché, cf.\ Tenreiro   \cite{Tenreiro}.
   
In the last quarter of the 20th century, the doctorates in Statistics of teachers from the University of Porto obtained abroad were those of Corália Vicente (Warwick, 1985), Margarida Brito (Paris VI, 1986), Paulo Gomes (Montpellier, 1987), Denisa Mendonça (Exeter, 1987), Carolina Silva (Exeter, 1989), Paula Brito (Paris IX, 1991), Eduarda Silva (Manchester, 1994), Joaquim Pinto da Costa (Rennes I, 1996), and Paulo Teles (Temple, 1999), who boosted the teaching and research of Statistics at ICBAS, Instituto de Ci\^{e}ncias Biom\'{e}dicas Abel  Salazar, the Faculty of Sciences, and the Faculty of Economics. Also from Porto, but from the Polytechnic Institute, Fernando Magalhães obtained his doctorate in Sheffield in 1997.

Meanwhile, between 1974 and 2000, the following teachers from the Technical University of Lisbon obtained their doctorates under the supervision of foreign scientists:
\begin{itemize}
 \item   Teachers from ISEG --- Daniel Muller (at ISEG but supervised by Marie Duflo, 1985), and abroad Lourdes Centeno (Heriot-Watt, Edinburgh, 1985), Nuno Crato (Delaware, 1992), and Alfredo Egídio dos Reis (Heriot-Watt, Edinburgh, 1994).
	\item  Teachers from IST --- Fernanda Ramalhoto (University College, 1977), João Branco (Newcastle, 1979), Acácio Porta Nova (Austin, 1985), João Amaral (Oxford, 1985), Daniel Paulino (São Paulo, 1989), António Pacheco (Cornell, 1994).
\item Teachers from ISA --- Carlos Agra Coelho (Michigan, 1992), Jorge Cadima (Kent, 1992).
 \end{itemize}
  
 During this period, the doctorates obtained abroad by teachers from the University of \'{E}vora were Carlos Braumann (Stony Brook, NY, 1979) and Russell Alpizar-Jara (North Carolina, 1997). From the University of Minho, Pedro Oliveira (Strathclyde, 1992), from the University of Algarve, Paulo Rodrigues (Manchester, 1999). From ISCTE, Manuela Magalhães Hill (Keele, 1987) and Elizabeth Reis (Southampton, 1998).

During this period, some foreigners decided to settle in Portugal: Kamil Feridun Turkman at FCUL; Ludwig Streit, first at the University of Minho and then at the University of Madeira, where he created the CCM -- Center for Mathematical Sciences, which is currently a pole of CIMA (Center for Research in Applied Mathematics) at \'{E}vora; Alpizar-Jara at the University of \'{E}vora; Manuel Scotto, who after obtaining his doctorate in Lisbon with Kamil Feridun Turkman, progressed at the University of Aveiro and later was appointed  Full Professor at IST; Giovanni Silva at IST; Laurens de Haan at CEAUL. Any of them  had an important impact on Statistics in Portugal. V. V. Yurinsky also settled in the University of Beira Interior but, apart from participating in a congress of the Portuguese Statistical Society, does not seem to have created a useful interrelation with Portuguese statisticians.
        
	Institutions like the Calouste Gulbenkian Foundation, FLAD, and NATO have been important funding and stimulating   research in Portugal. With the exchange over the past half-century, including ERASMUS, it would not be unreasonable to say that almost all of us are, to a greater or lesser extent, \textit{estrangeirados}\footnote{\, Or that only those who don't want to are not \textit{estrangeirados}. Note that there is currently a new breed of foreign-trained statisticians, promising young statisticians who, due to career progression constraints in Portugal, have decided to go abroad, hopefully only for a while. Notable among them are João Manuel Caravana Santos Silva (Essex), Pedro Miranda Afonso (Erasmus Medical Center, Rotterdam), Paulo Canas Rodrigues (Monash University, Melbourne), Miguel de Carvalho (Edinburgh), Lígia Henriques Rodrigues (São Paulo), Cláudia Neves (King's College), Nuno Sepúlveda (Warsaw), and Jorge Sinval (Singapore).}.
 
The FCT is also an important source of funding for internationalization, through scholarship and project programs, and through research centers. The creation of other research centers in Mathematics, with research lines in Probability, Stochastic Processes, and Statistics (such as the Center for Applied Mathematics (CIMA) at the University of \'{E}vora, with a second research line on Stochastic Processes, Statistics, and Operational Research, created in 1994, or the Center for Mathematics (CMAT) at the School of Sciences of the University of Minho (UM), with a branch at the University of Trás-os-Montes and Alto Douro (UTAD), with a research line in Statistics, Applied Probability, and Operational Research) had a huge effect on the internationalization of Portuguese research in Probability and Statistics, financing participation in numerous international events. The ``Bulletin of the Portuguese Statistical Society,'' Spring 2009, published information on several research centers (C.\ Agra Coelho, M.\ L.\ Esquível, and J.\ T.\ Mexia on CMA at the New University of Lisbon; M.\ Scotto on GPE at the University of Aveiro; C.\ A.\ Braumann and R.\ Alpizar-Jara on CIMA at the University of \'{E}vora; A.\ Pacheco on CEMAT at the Technical University of Lisbon; I.\ Fraga Alves on CEAUL).

  Portugal has hosted important international congresses, such as \vspace{0.35pc}
       
  Recent Advances in Statistics  \cite{Epstein}, Lisboa.\vspace{0.35pc}

  NATO ASI on Statistical Extremes and Application \cite{Tiago1984}, Vimeiro 1983.
\vspace{0.35pc}

 NATO ASI on Stochastic Analysis and Applications on Physics, Funchal 1993.
\vspace{0.35pc}

 Applied Stochastic Models and Data Analysis  (ASMDA 99), Lisboa 1999.
\vspace{0.35pc}

  23rd European Meeting of Statisticians \cite{Cunha}, Funchal 2001. 
\vspace{0.35pc}

32nd European Mathematical Psychology Group Meeting (EMPG 2001), Lisboa 2001.  \vspace{0.35pc}

VI International Congress on Insurance Mathematics and Econometry, 2002.  
\vspace{0.35pc}

IASC-IFCS Joint International Summer School on Classification and Data Mining in Business, Industry and Applied Research -- Methodological and Computational Issues (JISS 2003), Lisboa 2003.\vspace{0.35pc}

III International Conference on Extreme Value Analysis (EVA 2004) \cite{Hall},   Aveiro 2004.
\vspace{0.35pc}

 56th Session of the International Statistical Institute \cite{Gomes2007}, Lisboa 2007.
\vspace{0.35pc}

 Extremes in Vimeiro Today \cite{Fraga}, Vimeiro 2013.
\vspace{0.35pc}

 5th International Conference on Risk Assessment (ICRA5) \cite{Kitsos},  Tomar 2013.
\vspace{0.35pc}

Symbolic Data Analysis Workshop (SDA 2018) \cite{Brito}, Viana do Castelo 2018.
\vspace{0.35pc}

Workshop on New Frontiers in Statistics of Extremes (WNFSE 2020), Lisboa 2020.  
\vspace{0.35pc}

17th conference of the International Federation of Classification Societies (IFCS 2022), Porto 2022.
\vspace{0.35pc}

IMS International Conference on Statistics and Data Science (ICSDS) \cite{Xu}, Lisboa 2023.
\vspace{0.35pc}

These conferences brought a large number of foreign statisticians to Portugal, but there is no evidence to determine the effective added value this brought to the progress of Statistics in Portugal. The same can be said of the participation of foreign experts in the monitoring committees of research centres funded by the FCT. 

As far as visiting professors are concerned, David Mejzler's stay at FCUL inspired the work of Eugénia Graça Martins and Dinis Pestana   \cite{GMP} on classes of Mejzler, and that of W.\ Urfer was stimulating for several CEAUL researchers. Vic Barnett's brief stay at FCUL led to the SPRUCE project with Kamil Feridun Turkman, with the publication of
\bigskip

\hspace*{-1cm}\begin{minipage}{0.2\textwidth}
  \begin{itemize}
 \item []
\includegraphics[width=\linewidth]{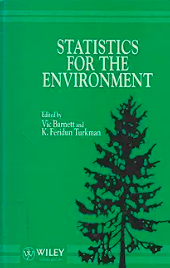}
\end{itemize}
\end{minipage}%
\begin{minipage}{0.6\textwidth}
\begin{itemize}
\item  []   Barnett, V.; Turkman, K.\ F. (1993). {\it Statistics for the Environment}, \cite{Barnett93}\\
Barnett, V.; Turkman, K.\ F. (1994). {\it Statistics for the Environment 2 --- Water Related Issues}, \cite{Barnett94}.\\
Barnett, V.; Turkman, K.\ F. (1997). {\it Statistics for the Environment   3 --- Pollution Assessment and Control}, \cite{Barnett97}.\\
Barnett, V.; Stein, A.; Turkman, K.\ F. (1998). {\it Statistics for the Environment 4 --- Statistical Aspects of Health and the Environment}, \cite{Barnett98}.\end{itemize}
\end{minipage}
\begin{minipage}{0.2\textwidth}
  \begin{itemize}
 \item []
\includegraphics[width=\linewidth]{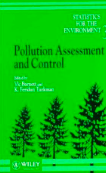}
\end{itemize}
\end{minipage}%
\bigskip

The University of Lisbon has awarded doctorates {\it Honoris Causa} to Vic Barnett, Ross Leadbetter, Laurens de Haan and Bento Murteira, in recognition of their contributions to the development of Statistics research in Portugal. The Universidade Nova de Lisboa awarded a doctorate {\it Honoris Causa} to C.\ R.\ Rao.

\bigskip

\subsubsection*{Scientific societies}

Once again, J.\ Tiago de Oliveira's dynamism led to the creation of the Portuguese Statistical and Operational Research Society (SPEIO), which in 1991 was transformed into the Portuguese Statistical Society (SPE), as the Portuguese Association for the Development of Operational Research had been created in the meantime.

The Portuguese Association for Data Classification and Analysis (CLAD) was also created in 1995.

The SPE regularly organises congresses, usually incorporating a mini-course, see the listing in Appendix 1.  Part of the proceedings of these congresses and the mini-courses are available at 
%\href{https://www.spestatistica.pt/pt}{https://www .spestatistica.pt/pt}
\url{https://www.spestatistica.pt/pt}. CLAD organises conferences (JOCLAD) with some proceedings available at 
%\href{https://clad.pt/}{https ://clad.pt/}
\url{https://clad.pt/}.

The SPE Congresses and JOCLAD are a barometer of statistics research in Portugal. For a co-ordinated overview of research up to 2005, see Murteira  (2005) \cite{Murt}.

The SPE elected Sir David Cox and C.\ R.\ Rao as foreign honorary members. Honorary members of SPE are also the Portuguese conferred  a Life Achievement Award, 
%\href{https://www.spestatistica.pt/en/socios/honorary-members}{https://www.spestatistica. pt/en/socios/honorary-members}
\url{https://www.spestatistica.pt/en/socios/honorary-members}.

Since 2006, Fernando Rosado has been doing invaluable work on what is happening in statistics teaching and research in Portugal, publishing an issue of the Bulletin of the Portuguese Statistical Society \cite{BolSPE} in the spring and autumn of each year, choosing a central theme for each issue. They are available at 
%\href{https://www.spestatistica.pt/publicacoes/categoria/boletim-da-spe}{https://www.spestatistica.pt/publicacoes/categoria/boletim-da-spe}
\url{https://www.spestatistica.pt/publicacoes/categoria/boletim-da-spe}.

The themes of all the issues that we owe to Fernando Rosado are: 

\vspace*{-.5cm}
\begin{center}
\begin{longtable}{cl}
Autumn 2006 & Teaching and Learning Statistics\\
Spring 2007 & The ``School of Extremes''\ in Portugal\\
Autumn 2007 & Biostatistics\\
Spring 2008 & ALEA --- A place of our world\\
Autumn 2008 & Stochastic Processes\\
Spring 2009 & Research (in) Statistics\\
Autumn 2009 & Econometric Models\\
Spring 2010 & Data Mining \\
Autumn 2010 & Spatial Statistics\\
Spring 2011 & Surveys and Censuses \\
Autumn 2011 & Survival Analysis\\
Spring 2012 & Statistics in Polytechnic Higher Education\\
Autumn 2012 & Statistical Methods in Medicine\\
Spring 2013 & Non-parametric Statistics\\
Autumn 2013 & The ``Bayesian School''\ in Portugal\\
Spring 2014 & (An) International Year of Statistics\\
Autumn 2014 & Statistics in Primary and Secondary Education\\
Spring 2015 & Statistics in Sport\\
Autumn 2015 & Statistics in Genetics\\
Spring 2016 & Time Series and their Applications\\
 Autumn 2016 & The Central Theme of Statistics\\
Spring 2017 & Uncertainty in Engineering\\
Autumn 2017 & The Central Theme of Statistics --- a new look\\
Spring 2018 & Multivariate Statistics --- perspective in the 21st century\\
Autumn 2018 & Stochastic Differential Equations and Some Applications\\
Spring 2019 & Integer Value Time Series\\
Autumn 2019 & Statistics in the Health Sciences\\
Spring 2020 & INE --- 85 years of statistics serving the country\\
Autumn 2020 & 40 years of SPE: Where did we come from? Where are we? Where are we going?\\
Spring 2021 & Covid Special: Statistics at the service of society\\
 Autumn 2021 & Machine Learning and Artificial Intelligence\\
Spring 2022 & Statistical Leadership\\
Autumn 2022 & Awards at the Portuguese Statistical Society\\
Spring 2023 & PORTSEA --- a Sea of Extremes in Portugal\\
Autumn 2023 &  Statistics (and) Education\\
Spring 2024 & Rising Stars
\end{longtable}
\end{center} 

\vspace*{-1cm}
The Spring 2008 issues, ALEA -- A Place in Our World, the Autumn 2014 issue, Statistics in Basic and Secondary Education, and the Autumn 2023 issue, Statistics (and) Education, address the important issue of pre-university education in Statistics. Sebastião e Silva was a pioneer regarding the teaching of Probability in high schools and had an influence on creating a branch for training Mathematics teachers at FCUL, for which he invited António Simões Neto\footnote{\, António Simões Neto was one of the first advocates of the Bayesian perspective. He published little, but see ``Coerência subjectivista e teoria ortodoxa das probabilidades'' \cite{Neto}.} to the teaching staff, a man of great culture, who influenced several generations of mathematicians (including the older co-author of this essay).

Jaime Carvalho e Silva  \cite{JCS}  reminds us of the interest of Bento de Jesus Caraça and Sebastião e Silva in this topic. More recently, António St. Aubyn, Eugénia Graça Martins, Jaime Carvalho e Silva, José Paulo Viana, João Branco, Luísa Canto e Castro Loura, and Manuela Neves, among others, have made significant contributions in this area. However, there have also been unfortunate interventions, and the issue is far from being stabilized (in fact, instability seems to be the hallmark of educational programs ...). Note the investment of INE, ALEA, and Explorística. Incidentally note that the Portuguese Society of Mathematics  has a sound investment in students and teachers' mathematical culture with the regular publication of the review {\it Gazeta de Matem\'{a}tica}:  In fact, one 
  of the objectives of the {\it Gazeta de Matemática}  is ``{\it to stimulate interest in the study of Mathematics, as well as the exchange of ideas among those who study, teach, research, use, or are simply interested in Mathematics}'', and consequently, it is a beneficial and stimulating read for innovation in education, including pre-university education. By accessing the Archive 
%\href{https://gazeta.spm.pt/arquivo}{https://gazeta.spm.pt/arquivo}
\url{https://gazeta.spm.pt/arquivo}, and searching for ``probability'' or ``statistics,'' one can immediately access the articles published in these areas.

Regarding the Spring 2011 issue, Surveys and Censuses, it should be noted that in 1991, the Center for Opinion Studies and Surveys (Universidade Católica) was created.

The Spring 2020 issue, INE -- 85 Years of Statistics Serving the Country, addresses the important topic of producing official statistics. For information on the National Statistical System, refer to the interesting book by
Ferreira da Cunha  \cite{FC}, and 
for more information on the INE,  other interesting reads are 
 Valério and  Bastien   \cite{VB} and the {\it Catálogo Bibliográfico} [Bibliographic Catalog] \cite{Catalog}.

Regarding the history of statistics in the specific context of official statistics, we  recommend 
Sousa's monograph  \cite{Hist}.

Besides its institutional role in producing and disseminating official statistics, INE has supported the progress of Statistics in Portugal and its internationalization. Notably, in 1996, INE launched the {\it Revista de Estatística} [Statistical Journal], which was internationalized in 2003 under the name REVSTAT. It can be accessed 
%\href{https://revstat.ine.pt/index.php/REVSTAT}{https://revstat.ine.pt/index.php/REVSTAT}
\url{https://revstat.ine.pt/index.php/REVSTAT}.

The Spring 2021 issue, Special COVID: Statistics at the Service of Society, reminds us of the interesting research in related areas, particularly Epidemiology. Manuel do Carmo Gomes, from the Department of Plant Biology at FCUL, has distinguished himself in this field. The initial phase of Manuel do Carmo Gomes' research was directed by José Manuel Campos Rosado, a Full Professor in the Genetics group at FCUL, who developed the area of Population Dynamics at FCUL.

The Spring 2023 issue focuses on PORTSEA -- a Sea of Extremes in Portugal, outlining the research in Extreme Value Theory. The acronym stands for \underline{PORT}uguese \underline{S}chool of \underline{E}xtremes and \underline{A}pplications. The acronym could have been \underline{PORT}uguese \underline{S}chool of \underline{EX}tremes (PORTSEX), which would have immediately implied its applications.

\subsubsection*{The Booksellers}

Statistics is an indispensable tool in a wide range of areas, and many courses now include at least one Statistics class. This has made the publication of books a profitable business, and currently, several publishers are producing what used to be lecture notes (this observation is not intended to be pejorative, it is simply a statement of fact). There is also a focus on publishing exercises with solutions and books illustrating data analysis with the help of statistical software. Notably, Maria Helena Gonçalves and Maria Salomé Cabral  \cite{GCA,GC} have  prepared interesting packages for the analysis of longitudinal data, and Camacho and Abreu \cite{CA} for Visual Survival Data Analysis.
\medskip

As we said at the beginning of this section, it is too early to have an idea of  what will survive from the multitude of  publications in this last half century. The selection criteria for those listed below was that they were useful reference books in various subjects, teaching cycles, or research, but some of them will be forgotten, since they are out of stock and unavailable.

\medskip

Murteira, B. (1988).
[{\it Statistics: Inference and Decision}]\cite{Murteira1988}.
\vspace{0.35pc}

Murteira, B.; Muller, D.;  Turkman, K.\ F. (1993).   [{\it  Analysis of Chronological Sequences}], \cite{MurteiraMullerTurkman1993}.
\vspace{0.35pc}

Tiago de Oliveira, J. (1990-91). [{\it Probability and Statistics --- Concepts, Methods and Applications}]\cite{TiagoOliveira1990}.
\vspace{0.35pc}

Guimarães, R. C. and Sarsfield Cabral, J.\ A. (1998).
 [{\it Statistics}] \cite{Guimaraes1998}.
\vspace{0.35pc}

Soares, A. (2000). [{\it Geoestatística para as Ciências da Terra e do Ambiente}] \cite{Soares2000}.
\vspace{0.35pc}

Pestana, D. and  Velosa, S. (2002). [{\it Introduction to Probability and Statistics}] \cite{PestanaVelosa2002}.
\vspace{0.35pc}

Paulino, C.\ D.;   Amaral Turkman, M.\ A.;   Murteira, B. (2003). [{\it Bayesian Statistics}],  \cite{Paulino2003}. (2nd edition in 2018, with G.\ Silva as co-author)
\vspace{0.35pc}

Gonçalves, E. and Mendes Lopes, N. (2003).
[{\it Statistics: Mathematical  Theory and Applications}] \cite{Mendes2003}.
\vspace{0.35pc}

Paulino, C.\ D. and  Singer, J.\ M. (2006).  [{\it Analysis of Categorical  Data}],  \cite{PaulinoSinger2006}.
\vspace{0.35pc}

 Suleman, A. (2009).      	
 [{\it Statistical Study of Fuzzy Sets]} \cite{Suleman2009}.
\vspace{0.35pc}

Murteira, B. and  Antunes, M. (2012).   [{\it Probability and Statistics}] \cite{MurteiraAntunes2012}.\bigskip

It is not very common for Portuguese researchers in this area to publish books in English, but we recommend:\medskip

Tiago de Oliveira, J. (1997).
{\it Statistical Analysis of Extremes} \cite{TiagoOliveira1997}.
\vspace{0.35pc}

de Haan, L.; Ferreira, A. (2006). {\it Extreme Value Theory: An Introduction} \cite{HaanFerreira2006}.
\vspace{0.35pc}

Turkman, K.\ F.; Scotto, M.; Bermudez, P.\ Z. (2014).
{\it Non-linear Time Series : Extreme Events and Integer Value Problems} \cite{TurkmanScottoBermudez2014}.
\vspace{0.35pc}

 Jacob, D; Neves, C.; Greetham, D.\ V. (2020). {\it Forecasting and Assessing Risk of Individual Electricity Peaks}. \cite{JacobNevesGreetham2020}.\vspace{0.35pc}

Morais, M.\ J.\ C. (2024).
{\it Stochastic Processes. Theory, Examples \& Exercises}
\cite{Morais2024}.

\bigskip

Some publishers continued to publish translations, for example:\medskip

Dagnelie, P. (1973--1975). {\it  Estatística, Teoria e Métodos}, translation of  {\it Théorie et Méthodes Statistiques: Applications Agronomiques}, \cite{Dagnelie1973}
\vspace{0.35pc}

Lévy, M.\ L. (1982). {\it Introdução à Estatística},  translation of {\it Comprendre les Statistiques},   \cite{Levy1982} 
\vspace{0.35pc}

Romano, R. (1989). {\it Enciclopédia Einaudi, vol 15 (Cálculo, Probabilidade)}, translation of {\it Einaudi Encyclopedia}, vol.\ 15 ({\it Probability, Calculus}), \cite{Romano1989}
(including the chapters   {\it Statistical Distribution}, {\it Probability}, and {\it Decision} by Bruno de Finetti, and a chapter by M.\ Mondadori on  {\it Statistical Induction})
  \vspace{0.35pc}
  
  Vassereau, A. (1989). {\it Introdução à Estatística}, translation of   [{\it La Statistique}],  \cite{Vassereau1989}.
  \vspace{0.35pc}
  
D'Hainaut, L. (1991 -- 1992). {\it Conceitos e Métodos da Estatística}, translation of {\it Concepts et M\'{e}thodes de la Statistique}, \cite{Hainaut1991}. \vspace{0.35pc}

Hoaglin, D.\ C.; Mosteller, F.; Tukey, J.\ W. (1992). {\it Análise Exploratória de Dados. Técnicas Robustas --- Um Guia}, translation of {\it Understanding Robust and Exploratory Data Analysis}, \cite{HoaglinMostellerTukey1992}. (It should be noted that he inspired Bento Murteira to rewrite his monograph on Descriptive Statistics, publishing {\it Análise Exploratória de Dados, Estatística Descritiva} [Exploratory Data Analysis, Descriptive Statistics], \cite{Murteira1993}.)
\vspace{0.35pc}

Mosteller, F.; Rourke, R.\ E.\ K. (1993).
{\it Estatísticas Firmes}, translation of {\it Sturdy Statistics}, \cite{MostellerRourke1993}.\vspace{0.35pc}

Ghiglione, R.; Matalon, B. (1993).
{\it O Inquérito --- Teoria e Prática}, translation of {\it Sociological Surveys: Theory and Practice}, \cite{GhiglioneMatalon1993}.
\bigskip

60 years late, the famous  {\it How to Lie with Statistics}, by D. Huff was translated\footnote{\, Recognizing the impact of this book, the magazine {\it Statistical Science} dedicated in 2005 part of {\bf 20} (3) to a ``SPECIAL SECTION: HOW TO LIE WITH STATISTICS TURNS FIFTY'', with articles by J.\ M.\ Steele ``Darrell Huff and Fifty Years of How to Lie with Statistics'', J. \ Best ``Lies, Calculations and Constructions: Beyond How to Lie with Statistics'', M.\ Monmonier ``Lying with Maps'', W.\ Kr\"{a}mer and G.\ Gigerenzer ``How to Confuse with Statistics or: The Use and Misuse of Conditional Probabilities'', R.\ D.\ De Veaux and D.\ J.\ Hand ``How to Lie with Bad Data'', C.\ Murray ``How to Accuse the Other Guy of Lying with Statistics'', S.\ C.\ Morton ``Ephedra'', S.\ E.\ Fienberg and P.\ C.\ Stern ``In Search of the Magic Lasso: The Truth About the Polygraph'', 
%\href{https://projecteuclid.org/journals/statistical-science/volume-20/issue-3}{https://projecteuclid.org/journals/statistical-science/volume-20/issue-3}
\url{https://projecteuclid.org/journals/statistical-science/volume-20/issue-3}.
}. In fact, currently several publishers are publishing lighter entertaining  books, for example: \medskip

Clegg, F. (1995). {\it Estatística para Todos: um Manual para Ciências Sociais}, translation of {\it Simple Statistics: a Course Book for the Social Sciences}, \cite{Clegg1995}.\vspace{0.35pc}

Marques de Sá, J. (2006). {\it  O Acaso --- A Vida do Jogo e o Jogo da Vida}  [Chance: The Life of Games and the Game of Life],  \cite{MarquesSa2006}
(very interesting, we hope that in the reissue the typos will be corrected on p.\ 33, lines 4--6, where it should be 0.35 instead of 0.45, and on
p.\ 58 in the table the premium case is missing behind 1, ``initial choice 1, presenter shows door 3, if you change you lose'').
\vspace{0.35pc}

Mlodinow, L. (2009). {\it O Passeio do Bêbado: como o Acaso Rege as nossas Vidas}, translation of {\it The Drunkard's Walk: How Randomness Rules our Lives}, \cite{Mlodinow2009}.
\vspace{0.35pc}

Grima, P. (2011). {\it A Certeza Absoluta e  Outras Ficções: os Segredos da Estatística} , translation of {\it La Certitude Absolue et Autres Illusions : Les Secrets de la Statistique}, \cite{Grima2011}.\vspace{0.35pc}

Huff, D. (2013). {\it  Como Mentir com Estatística}, translation of {\it How To Lie With Statistics}, \cite{Huff2013}.
\vspace{0.35pc}

Blauw, S. (2020). {\it  O Poder dos Números: Como Estatísticas, Percentagens e Análises nos Enganam e Desenganam}, translation of {\it The Number Bias: How Numbers Lead and Mislead Us}, \cite{Blauw2020}.

\section{Final considerations}

Although there is no translation of {\it Statistical Methods for Research Workers} published in Portugal, as far as we were able to ascertain, Fisher's influence was notable on teachers and researchers at the Institute of Anthropology of Coimbra and the Instituto Superior de Agronomia of Lisbon, which introduced in Portugal notions of Analysis of Variance and Experiment Planning, properties of estimators and significance tests.

The Faculty of Law of the University of Coimbra, where Political Economy included elementary notions of Statistics, was more involved with Descriptive Statistics and official statistics. Mathematical Statistics only appeared timidly with the notes by M. Santos and Zaluar Nunes, later developed by Leite Pinto and especially by Bento Murteira.

The Faculties of Science and the Higher Institutes of the Technical University of Lisbon were pivotal in the teaching of Probability, Errors, and Statistics, although, with some justification,  Fernandes Costa \cite{Fernandes} lamented ``{\it the name given to the new discipline, stigmatizing it from birth}''. Pedro Braumann, after initial training as an assistant in this subject at FCUL, became the researcher with the most advanced works in Probability Theory following his internship at Stanford. More applied areas, such as Quality Control and Reliability or Queueing Theory, have significant prominence, particularly in Engineering Institutes. Biostatistics, Survival Analysis, Clinical Trials, and Meta-Analysis are increasingly important in Life Sciences applications. In addition to the Faculties of Science and the Institute of Agronomy, noteworthy groups in this field include those at the University of Minho and the Abel Salazar Institute of Biomedical Sciences, the Faculties of Medicine, and the National School of Public Health. 

Spatial Statistics has also seen significant development,  
 particularly at FCUL and IST of the University of Lisbon, at FCT and IMS of the New University of Lisbon, at the University of \'{E}vora, and at the University of Minho. The Open University has invested in the area of Experimental Design, and the Catholic University, as previously mentioned, in Sampling. Both the University of Porto and the University of Coimbra have internationally renowned groups in the aforementioned areas and beyond, such as Stochastic Processes,   Time Series, and Densities Estimation, spread across their various institutions. Environment Statistics is also a field in development in various universities, and Population Dynamics and Epidemics, for instance in the University of Lisbon, or Fisheries Stocks (\'{E}vora, Algarve), and several areas of Stochastic Analysis and of Stochastic Differential Equations (\'{E}vora, Lisboa). 
 
 The works of  M.\ Ivette Gomes on the History of Statistics highlights that Extreme Value Theory refers  important researchers in this area  at Universities and Polytechnic Institutes in Portugal \cite{IGomes}. In this brief overview, undoubtedly filled with unjust omissions, we only mention the institutions where Statistics groups have reached a critical mass to stand out, demonstrating the need for investment to reward the work already done at the Universities of Aveiro, Algarve, Azores, Madeira, Beira Interior, and UTAD, as well as at the Polytechnic Institutes, allowing them to gain visibility. In any case, comparing the current situation, both in terms of the number of faculty/researchers and the educational offerings, with what was happening fifty years ago, it is evident that we must thank our mentors, who paved the initial way, and commend the visionary spirit of Veiga Simão, who profoundly transformed higher education in Portugal. 

After the brief foray by Daniel Augusto da Silva into Actuarial Calculus, mention should be made of Arthur Malheiros' book    \cite{Malheiros}, available at the Insurance and Pension Funds Supervisory Authority.  Regarding the teaching and research in Actuarial Calculus, refer to Lourdes Centeno  \cite{Centeno} for the history of its teaching in Portuguese universities.

It is noteworthy that Statistics continues to attract scientists from diverse backgrounds (for example, Carlos Daniel Paulino started as a chemical engineer, Tiago Marques and Manuel Carmo Gomes are biologists, and J. M. Campos Rosado was initially a veterinarian), and it develops in various institutions --- and we should not forget the Bank of Portugal --- which definitively contributes to its versatility.

Varennes e Mendonça was not the only distinguished professor from other fields who, by necessity, accepted the chair of Probability. To our knowledge, and certainly missing others, Pedro Lago at the Faculty of Sciences of the University of Porto and Manuel Neto Murta in Coimbra also chaired Probability courses. The Mathematics Department of the University of Coimbra, honoring the good tradition of settling its debts, organized the 1989 Probability and Statistics Days in honor of Professor Manuel Neto Murta, publishing {\it Estudos de Probabilidades e Estatística}  (Probability and Statistics Studies) with the works presented by C. Gourieroux and I. Peaucelle, R. Moché, M. I. Gomes, E. Gonçalves, M. Delecroix, M. N. Mendes Lopes, and J. Tiago de Oliveira.

Sampling and Experimental Design are still not fundamental courses for good training in scientific research methodology at many higher education institutions. The 2001 book by   Vicente {\it et al.} \cite{Vicente} is an obvious recommendation for anyone wanting to learn the most common sampling techniques, as is the SPE mini-course by Paulo Gomes, {\it Tópicos de Sondagens} [Survey Topics], available at 
%\href{https://spestatistica.pt/publicacoes/publicacao/topicos-de-sondagens}{https://spestatistica.pt/publicacoes/publicacao/topicos-de-sondagens}
\url{https://spestatistica.pt/publicacoes/publicacao/topicos-de-sondagens}. The famous book by  Kinsey {\it et al.} \cite{Kinsey} (Portuguese translation published  by Meridiano, but to our knowledge out of print) has a remarkable chapter on Sampling, with many judicious observations on, for example, the art of interviewing. Incidentally, we recall that Kinsey's book provoked great controversy, leading the American Association of Statistics to request renowned statisticians to publish a report evaluating the quality of the data collection and analysis. The result, Cochran {\it et al.} \cite{Cochran} is a notable book that we recommend for developing the critical thinking of all Statistics users.

With the growth of human resources, universities have created departments, Master's and Doctoral programs in Probability and Statistics and related areas. Notably, the Institute of Statistics and Information Management (ISEGI) was established in 1989 at the New University of Lisbon, with strong ties to INE, later renamed NOVA IMS -- Information Management School, and the degree in Applied Mathematics to Economics and Management (MAEG) at ISEG/UTL. The Francisco Manuel dos Santos Foundation, established in 2009, has also played a significant role in disseminating statistical data about Portugal through PORDATA (\url{https://www.pordata.pt/portugal}).

The preparation of this work led us to realize how inaccessible works of our scientific heritage are. It would certainly be reasonable to digitize these ``reserved'' works and eventually make them accessible on the SPE website, or even suggest to the FCT or the Academy of Sciences to support a project to create a digital library that would allow access to those interested in the History of Science.

Additionally, some books published in the last quarter-century are unavailable because the Calouste Gulbenkian Foundation holds the copyright and does not reissue them (for instance the works of Sebastião e Silva \cite{SebastiaoSilva1999} and the book by Pestana and Velosa \cite{PestanaVelosa2002}, and because Livraria Escolar Editora has disappeared (namely the books by Esmeralda Gonçalves and Nazaré Mendes Lopes \cite{Mendes2003}, and by Bento Murteira and Marília Antunes \cite{MurteiraAntunes2012}). Surely an institution like SPE or the centers of those authors could reach an agreement for the rights to reprint, at least in PDF.\bigskip

\noindent\begin{minipage}{0.18\textwidth}
  \begin{itemize}
 \item []
\includegraphics[width=\linewidth]{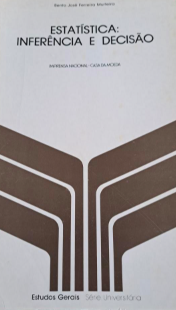}
\end{itemize}
\end{minipage}%
\hfill%
\begin{minipage}{0.75\textwidth}
\begin{itemize}
\item  [] ``Bento Murteira's more daring  work. {\it Estatística: Inferência e Decisão} [Statistics: Inference and Decision],   published in 1988, due to the deep critical reflection on the statistical methodologies it constitutes, was in our midst a milestone that promoted a critical awareness of statistical theory and practice that, even today, must be a regular and reflective read for those who make Statistics their professional occupation.''  --- Daniel Paulino  (2005) \cite{Paulino}.  \end{itemize}
\end{minipage}\\

\medskip

Once again, a call for republication: surely accredited institutions could insist with Imprensa Nacional -- Casa da Moeda to reprint, or possibly transfer the rights to an institution that would do so, since the 2024  centenary reissue is limited to 100 books intended to be offered to libraries.

Finally, to lighten so much scientific bibliography: references to Probability and Statistics in literature are an interesting indicator of the social recognition of these disciplines. Already in the 19th century, Sherlock Holmes praises Probability:
 \begin{quotation}
``{\it We balance probabilities and choose the most likely. It is the scientific use of the imagination.}'' ~ Arthur Conan Doyle, {\it The Hound of the Baskervilles.}
\end{quotation}
Meanwhile, Mark Twain (who also contributed with a ``data analysis'' of sorts, ``{\it quitting smoking is easy; I've done it over a hundred times}'') attributes --- it is a lie --- the aphorism ``{\it There are three kinds of lies: lies, damned lies, and statistics}'' to Disraeli.

Isaac Asimov's {\it  Foundation}, reputedly the most famous science fiction series, is driven by ``psychohistory'', a new branch of Probability, and in particular, the second volume, {\it Foundation and Empire}, extensively discusses risk and the management of low probabilities. It ensures a pleasurable read.

{\it An Instance of the Fingerpost} by Ian Pears, translated into Portuguese as ``O Círculo da Cruz'' (Livros do Brasil), is a fascinating detective novel with a very interesting description of the invention of the experimental method, foreshadowing Fisher's experimental design.

  {\it A   Conspiracy of Paper} by David Liss features a hero, Benjamin Weaver, a Jew of Portuguese descent who becomes embroiled in the financial fraud of the South Sea Company. Being an amateur probabilist, the book contains a set of interesting information about intellectuals who dealt with the early developments of Probability in the United Kingdom, less famous than the exiled Frenchman Abraham de Moivre. A fascinating read.
 
\bigskip

 \section*{Appendix 1}

\noindent {\bf Short Courses, Congresses of the Portuguese Statistical Society, downloadable at 
%\href{https://www.spestatistica.pt/en/publications/category/short-course}{https:// www.spestatistica.pt/en/publications/category/short-course}
\url{https://www.spestatistica.pt/en/publications/category/short-course}
}
\bigskip

 Gomes, P. (1998)
{\it Tópicos de Sondagens} [Survey Topics]
\vspace{0.35pc}

 Gomes, M.\ I.; Figueiredo, F.; Barão, M.\ I. (1999)  2\textordfeminine ed.\ 2010,
{\it Controlo Estatístico da Qualidade} [Statistical Quality Control] 
\vspace{0.35pc}

  Amaral Turkman, M. A.; Silva, G. (2000)
{\it Modelos Lineares Generalizados --- da Teoria à Prática} [Generalized Linear Models --- from Theory to Practice]
\vspace{0.35pc}

  Brilhante, M.\ F.; Pestana, D.; Rocha, J.; Rocha, M.\ L.; Velosa, S. (2001)  2ª ed. 2011,
{\it Inferência Estatística sobre a Localização Usando a Escala} [Statistical Inference about Location Using the Scale]
\vspace{0.35pc}

  Muller, D. (2002)
{\it Modelos Heterocedásticos. Aplicações com o software Eviews} [Heteroscedastic Models. Applications with Eviews software]
\vspace{0.35pc}

  Gonçalves, E.; Mendes Lopes, N. (2003)  2\textordfeminine ed.\ 2008
{\it Séries Temporais: Modelações Lineares e não Lineares} [Time Series: Linear and Non-Linear Modeling]
\vspace{0.35pc}

  Branco, J.\ A. (2004)
{\it Uma Introdução à Análise de Clusters} [An Introduction to Cluster Analysis]
\vspace{0.35pc}

  Braumann, C. (2005)
{\it Introdução às Equações Diferenciais Estocásticas e Aplicações} [Introduction to Stochastic Differential Equations and Applications]
\vspace{0.35pc}

  Rosado, F. (2006)
{\it Outliers em Dados Estatísticos} [Outliers in Statistical Data]
\vspace{0.35pc}

  Pires , A.\ M.; Branco, J.\ A. (2007)
{\it Introdução aos Métodos Estatísticos Robustos} [Introduction to Robust Statistical Methods]
\vspace{0.35pc}

  Carvalho, M.\ L.; Natário, I.\ C. (2008)
{\it Análise de Dados Espaciais} [Spatial Data Analysis]
\vspace{0.35pc}

  Rocha, C.; Papoila, A.\ L. (2009)
{\it Análise de Sobrevivência} [Survival Analysis]
\vspace{0.35pc}

  Tenreiro, C. (2010)
{\it Uma Introdução à Estimação Não-Paramétrica da Densidade} [An Introduction to Non-Parametric Density Estimation]
\vspace{0.35pc}

  Cabral, M.\ S.; Gonçalves, M.\ H. (2011)
{\it Análise de Dados Longitudinais} [Longitudinal Data Analysis]
\vspace{0.35pc}

  Salgueiro, M.\ F. (2012)
{\it Modelos com Equações Estruturais} [Models with Structural Equations]
\vspace{0.35pc}

  M. Ivette Gomes, M.\ I.;   Fraga Alves, M.\ I., Neves, C. (2013)
{\it Análise de Valores Extremos: uma Introdução} [Extreme Value Analysis: an Introduction]
\vspace{0.35pc}

   Amaral Turkman, M.\ A.; Paulino, C.\ D. (2015)
{\it Estatística Bayesiana Computacional --- uma Introdução} [Computational Bayesian Statistics --- an Introduction]
\vspace{0.35pc}

  Brilhante, M.\ F. (2017)
{\it Uma Introdu\c{c}\~{a}o \`{a} Meta-An\'{a}lise} [An Introduction to Meta-Analysis]
\vspace{0.35pc}

  Amado, C.; Nunes,  C.; Sardinha, A. (2019)
{\it Análise Estatística de Dados Financeiros} [Statistical Analysis of Finance Data]
\vspace{0.35pc}

  Afonso, P.\ M. (2023).   {\it Modelação Conjunta de Dados Longitudinais e de Sobrevivência} [Joint Modeling of Longitudinal and Survival Data]
  \bigskip

Before the SPE Congresses, CEAUL organized conferences, whose proceedings were published:
\smallskip 

 {\it Actas,  I Colóquio de Estatística e  Investigação  Operacional}, Lisboa, Centro de Estatística e Aplicações, 1977. 
 \vspace{0.35pc}
 
{\it Actas, II Colóquio Estatística e  Investigação  Operacional}, Fundão/Covilhã, CEAUL/ SPEIO,  1981.\vspace{0.35pc}
 
 {\it Actas, III Colóquio Estatística e  Investigação  Operacional}, CEAUL/SPEIO, Tróia, 1985.
 \medskip

Proceedings of the SPE Congresses:
\medskip
		
	{\it A Estatística e o Futuro e o Futuro da Estatística} [Statistics and the Future and the Future of Statistics], Actas do I Congresso da SPE, 1993, Vimeiro; ed.\  Dinis Pestana, Antónia Turkman, João Branco, Luísa Canto e Castro e Ana Pires.
   \vspace{0.35pc}
   
  {\it Actas do II Congresso da SPE}, 1994, Luso; ed.\  M. Nazaré Mendes Lopes, M. Esmeralda Gonçalves, M. Emília Nogueira, Ana Cristina Rosa e Helena Ferreira.
  \vspace{0.35pc}
  
{\it Bom Senso e Sensibilidade: Traves Mestras da Estatística} [Common Sense and Sensitivity: Main Supports of Statistics], Actas do III Congresso da SPE, 1995, Guimarães; ed.\  João Branco, Paulo Gomes e Jorge Prata.
 \vspace{0.35pc}
 
{\it A Estatística a Decifrar o Mundo} [Statistics Deciphering the World], Actas do IV Congresso da SPE, 1996, Funchal; ed.\  Rita Vasconcelos, Isabel Fraga Alves, Luisa Canto e Castro e Dinis Pestana.
 \vspace{0.35pc}
 	
{\it Estatística -- a Diversidade na Unidade} [Statistics -- Diversity in Unity], Actas do V Congresso Anual da Sociedade Portuguesa de Estatística, 1997, Curia; ed.\  Manuela Souto de Miranda e Isabel Pereira.
 \vspace{0.35pc}
 	
{\it Afirmar a Estatística: um Desafio para o Século XXI} [Asserting Statistics: a Challenge for the 21st Century], Atas do VI Congresso da SPE, 1999, Tomar; ed.\  Carlos Daniel Paulino, António Pacheco, Ana Pires e Ferreira da Cunha.
 \vspace{0.35pc}
 
{\it Um Olhar sobre a Estatística} [A Look upon Statistics], Actas do VII Congresso da SPE, 2001, Ofir; ed.\  Pedro Oliveira e Emília Athayde.
 \vspace{0.35pc}
 
{\it A Estatística em Movimento} [Statistics in Motion], Actas do VIII Congresso da SPE, 2001, Peniche; ed.\  Maria Manuela Neves, Jorge Cadima, Maria João Martins e Fernando Rosado.
 \vspace{0.35pc}
 		
{\it Novos Rumos em  Estatística} [New Directions in Statistics], Actas do IX Congresso da SPE, 2002, Ponta Delgada; ed.\  Lucília Carvalho, Fátima Brilhante e Fernando Rosado.
 \vspace{0.35pc}
 
{\it Literacia e Estatística} [Literacy and Statistics], Actas do X Congresso da SPE, 2003, Porto; ed.\  Paula Brito, Adelaide Figueiredo, Fernanda Sousa, Paulo Teles e Fernando Rosado.
 \vspace{0.35pc}
 		
{\it Estatística com Acaso e Necessidade} [Statistics with Chance and Necessity], Actas do XI Congresso da SPE, 2004, Faro; ed.\  Paulo M. M. Rodrigues, Efigénio da Luz Rebelo e Fernando Rosado.
 \vspace{0.35pc}
 
{\it Estatística Jubilar} [Jubilar Statistics], Actas do XII Congresso Anual da SPE, 2005, \'{E}vora; ed.\  Carlos Braumann, Paulo Infante, Manuela M. Oliveira, Russell Alpízar-Jara e  Fernando Rosado.
 \vspace{0.35pc}
 
{\it A Ciência Estatística} [Statistical Science], Actas do XIII Congresso da SPE, 2006, Ericeira; ed.\  Luísa Canto e Castro, Eugénia Graça Martins, Cristina Rocha, M. Fernanda Oliveira, Margarida Mendes Leal e Fernando Rosado.
	 \vspace{0.35pc}
	 
{\it Estatística,  Ciência Interdisciplinar} [Statistics, an Interdisciplinary Science], Actas do XIV Congresso da SPE, 2007, Covilhã; ed.\  Maria Eugénia Ferrão, Célia Nunes e Carlos Braumann.
 \vspace{0.35pc}

{\it Estatística: da Teoria à Prática} [Statistics: from Theory to Practice], Actas do XV Congresso da SPE, 2008, Lisboa; ed.\ Manuela Magalhães Hill, Manuel Alberto Ferreira, José G. Dias, Maria de Fátima Salgueiro, Helena Carvalho, Paula Vicente e Carlos Braumann.
 \vspace{0.35pc}
 
{\it Estatística: Arte de Explicar o Acaso} [Statistics: the Art of Explaining Chance], Actas do XVI Congresso Anual da SPE, 2009, Vila Real; ed.\ Irene Oliveira, Elisabete Correia, Fátima Ferreira, Sandra Dias e Carlos Braumann.
 \vspace{0.35pc}
 
{\it Advances in Regression, Survival Analysis, Extreme Values, Markov Processes and Other Statistical Applications}, Selected Papers of the Statistical Societies, Springer, XVII Congresso Anual da SPE, 2013, Sesimbra; Ed.\ João Lita da Silva, Frederico Caeiro, Isabel Natário, Carlos A. Braumann, Manuel L. Esquível and João Tiago Mexia.  
 \vspace{0.35pc}
 
{\it Recent Developments in Modeling and Applications in Statistics}, Selected Papers of the Statistical Societies, Springer, XVIII Congresso Anual da SPE, 2013, São Pedro do Sul; ed,\ Paulo Eduardo Oliveira, Maria da Graça Temido, Carla Henriques e Maurizio Vichi.
 \vspace{0.35pc}
 	
{\it New Advances in Statistical Modeling and Applications}, Selected Papers of the Statistical Societies, Springer, XIX Congresso Anual da SPE, 2014, Nazaré; ed.\  António Pacheco, Rui Santos, Maria do Rosário Oliveira e Carlos Daniel Paulino.
\vspace{0.35pc}

{\it Estatística: Novos Desenvolvimentos e Inspirações} [Statistics: New Developments and Inspirations], Actas do XX Congresso Anual da SPE, 2013, Porto; ed.\ Manuela Maia, Pedro Campos e Pedro Duarte Silva.
 \vspace{0.35pc}

{\it Estatística: a ciência da incerteza} [Statistics: the science of uncertainty], Atas do XXI Congresso da SPE, 2014, Aveiro; ed.\ Isabel Pereira, Adelaide Freitas, Manuel Scotto, Maria Eduarda Silva e Carlos Daniel Paulino. 
%\href{https://www.spestatistica.pt/publicacoes/publicacao/estatIstica-ciencia-incerteza}{https://www.spestatistica.pt/publicacoes/publicacao/estatIstica-ciencia-incerteza}
\url{https://www.spestatistica.pt/publicacoes/publicacao/estatIstica-ciencia-incerteza}
\vspace{0.35pc}

{\it Estatística: progressos e aplicações} [Statistics: Progresses and Applications], Atas do XXII Congresso da SPE, 2015, Olhão; ed.\ Clara Cordeiro, Conceição Ribeiro, Carlos Sousa, Maria Helena Gonçalves, Nelson Antunes e Maria Eduarda Silva. 
%\href{https://www.spestatistica.pt/publicacoes/publicacao/estatIstica-progressos-aplicacoes}{https://www.spestatistica.pt/publicacoes/ publicacao/estatIstica-progressos-aplicacoes}
\url{https://www.spestatistica.pt/publicacoes/publicacao/estatIstica-progressos-aplicacoes}
\vspace{0.35pc}

{\it Atas do XXIII Congresso da SPE} [Proceedings of the XXIII SPE Congress], 2017, Lisboa; ed.\ Maria de Fátima Salgueiro, Paula Vicente, Teresa Calapez, Catarina Marques e Maria Eduarda Silva.  
%\href{https://www.spestatistica.pt/publicacoes/publicacao/atas-do-xxiii-congresso-da-spe}{https://www.spestatistica.pt/publicacoes/publicacao/atas-do-xxiii-congresso-da-spe}
\url{https://www.spestatistica.pt/publicacoes/publicacao/atas-do-xxiii-congresso-da-spe}
\vspace{0.35pc}

{\it Estatística: Desafios Transversais às Ciências com Dados} [Transversal challenges to Sciences with Data], Atas do XXIV Congresso da Sociedade Portuguesa de Estatística, 2021, Amarante ed.\ Paula Milheiro, António Pacheco, Bruno de Sousa, Isabel Fraga Alves, Isabel Pereira, Maria João Polidoro e Sandra Ramos. 
\url{https://www.spestatistica.pt/publicacoes/publicacao/estatistica-desafios-transversais-ciencias-com-dados}\vspace{0.35pc}

{\it Recent Developments in Statistics and Data Science}, Atas do XXV Congresso da Sociedade Portuguesa de Estatística, 2021, \'{E}vora (ed. Regina Bispo, Lígia Henriques-Rodrigues, Russell Alpizar-Jara e Miguel de Carvalho). 
\url{https://www.spestatistica.pt/publicacoes/publicacao/recent-developments-statistics-and-data-science}
%\href{https://www.spestatistica.pt/publicacoes/publicacao/recent-developments-statistics-and-data-science}{https://www.spestatistica.pt/publicacoes/publicacao/recent-developments-statistics-and-data-science}
\vspace{0.35pc}

{\it New Frontiers in Statistics and Data Science}, Springer Proceedings in Mathematics \& Statistics, Springer, XXVI Congresso da SPE, 2024, Guimarães (ed.\ Lígia Henriques-Rodrigues, Raquel Menezes, Luís Meira Machado, Miguel de Carvalho e Susana Faria).

\newpage
\section*{Appendix 2}
Table of Contents  of the {\it Manual de Estatística Médica} \cite{ReisSarmento1960} [{\it Manual of Medical Statistics}] by Carlos Santos Reis and Alexandre Sarmento.
\bigskip

{\footnotesize
\begin{tabular*}{0.85\textwidth}{@{\extracolsep{\fill}} lr } 
Introduction & 3\\[3pt]
1 -- Fundamental concepts & 6\\[3pt]
2 -- Statistical methods & 14\\
~~~1 -- Definition & 14\\
~~~2 -- Objectives & 14\\
~~~3 -- Purpose & 14\\
~~~4 -- Feature & 14\\
~~~5 -- Observation methods & 14\\
~~~6 -- Applications & 14\\
~~~7 -- Limitations & 14\\[3pt]
\multicolumn{2}{c}{\textsc{First part}} \\[3pt]
Application of the statistical method & 25\\[3pt]
3 -- Planning & 26\\[3pt]
4 -- Organization & 28\\[3pt]
5 -- Preparation & 31\\
~~~1 -- The collection or collection of data & 32\\
~~~~~~1.1 -- Collection method & 33\\
~~~~~~1.2 -- Observation methods & 34\\
~~~~~~1.3 -- Scope of investigations & 36\\
~~~~~~1.4 -- Investigation modalities & 37\\
~~~~~~1.5 -- Sources of data & 42\\
~~~~~~1.6 -- Preparation of questionnaires & 43\\
~~~~~~1.7 -- Pilot survey & 46\\
~~~~~~1.8 -- Sources of errors in surveys & 48\\
~~~~~~1.9 -- Observations record & 52\\[3pt]
6 -- Data verification & 56\\
~~~1 -- Critique of results & 56\\
~~~2 -- Data ordering & 58\\
~~~3 -- Data classification & 61\\
~~~4 -- Organization of a frequency distribution & 64\\
~~~5 -- Enumeration of individuals & 68\\
~~~6 -- Coding of results & 75\\[3pt]
7 -- Presentation of data & 81\\
~~~1 -- Text presentation & 81\\
~~~2 -- Tabular presentation & 82\\[3pt]
8 -- Graphical presentation & 92\\
~~~1 -- General considerations & 92\\
~~~2 -- Main fundamentals and criticisms of graphic presentation & 93\\
~~~3 -- Rules for constructing graphs & 93\\
~~~4 -- Definition and major divisions in graphic presentation & 95\\[3pt]
\end{tabular*}

\newpage
\begin{tabular*}{0.85\textwidth}{@{\extracolsep{\fill}} lr } 
9 -- Use & 113\\
~~~1 -- Absolute values & 114\\
~~~2 -- Derived values & 114\\[3pt]
10 -- Description of statistical series & 124\\
~~~1 -- Qualitative characters & 125\\
~~~2 -- Quantitative characters & 126\\[3pt]
11 -- Location measurements & 129\\
~~~1 -- Median & 130\\
~~~2 -- Arithmetic mean & 133\\
~~~3 -- Mode & 137\\
~~~4 -- Geometric mean & 138\\
~~~5 -- Comparison of the main location measurements & 142\\[3pt]
12 -- Dispersion and shape measurements & 145\\
~~~1 -- Dispersion measurements & 145\\
~~~2 -- Shape measurements & 152\\
~~~3 -- Presentation of results & 153\\[3pt]
13 Simplified calculation of parameters & 154\\
~~~1 -- Mode & 155\\
~~~2 -- Median & 156\\
~~~3 -- Arithmetic mean & 157\\
~~~4 -- Variance & 160\\
~~~5 -- Joint calculation of mean and variance & 164\\[3pt]
14 -- Interpretation of data. Statistical schemes & 167\\[3pt]
15 -- Some theoretical distributions & 174\\
~~~1 -- Notion of probability and law of large numbers & 175\\
~~~2 -- Binomial distribution & 178\\
~~~3 -- Normal distribution & 182\\
~~~4 -- Poisson distribution & 189\\
~~~5 -- Variable transformation & 191\\[3pt]
16 -- Sampling & 193\\
~~~1 -- Sampling deformation & 194\\
~~~2 -- Confidence limits & 196\\
~~~3 -- Sample representativeness & 196\\
~~~4 -- Sampling types & 197\\
~~~5 -- Number of observations & 202\\
~~~6 -- Sampling problems & 204\\[3pt]
17 -- Estimation and confidence limits & 205\\
~~~1 -- Estimation and confidence limits of the mean & 207\\
~~~2 -- Estimation and confidence limits of the  variance & 210\\
~~~3 -- Estimation and confidence limits of a proportion & 210\\[3pt]
18 -- Compliance tests & 215\\
~~~1 -- Conformity tests for parameters & 219\\
~~~2 -- Distribution of conformance tests & 220\\[3pt]
19 -- Homogeneity problem & 232\\
~~~1 -- Comparison of two means & 234\\
~~~2 -- Comparison of two variances or two standard deviations & 245\\
~~~3 -- Comparison of two rates or proportions & 248\\[3pt]
\end{tabular*}

\begin{tabular*}{0.85\textwidth}{@{\extracolsep{\fill}} lr }
\multicolumn{2}{c}{\textsc{Appendix}} \\[3pt]
Calculation of factorials & 257\\[3pt]
20 -- Comparison of multiple samples & 263\\
~~~1 -- Simultaneous comparison of multiple means & 265\\
~~~2 -- Simultaneous comparison of several variances & 304\\
~~~3 -- Simultaneous comparison of several proportions & 306\\[3pt]
21 -- Statistical dependence & 309\\
~~I -- Relationships between two qualitative characters. Notions of independence and association & 309\\
~~~1 -- Presentation of results & 310\\
~~~2 -- Notion of $\ll$independence$\gg$ and $\ll$association$\gg$ & 311\\
~~~3 -- Association issues & 312\\
~~~4 -- Existence of the association & 312\\
~~~5 -- Independence tests & 314\\
~~~6 -- Different types of problems & 321\\
~~~7 -- Association measures & 329\\[3pt]
22 -- Statistical dependence & 332\\
~~II -- Relationships between two quantitative characters. Understanding correlation and regression & 332\\
~~~1 -- Understanding correlation and regression & 333\\
~~~2 -- Correlation types & 335\\
~~~3 -- Presentation of results & 336\\
~~~4 -- Correlation form & 339\\
~~~5 -- Regression line & 342\\
~~~6 -- Regression standard error & 350\\
~~~7 -- Correlation coefficient & 351\\
~~~8 -- Numerical calculation of correlation parameters & 353\\
~~~9 -- Interpretation of correlation coefficients & 364\\
~~~10 -- Significance of a correlation coefficient & 366\\
~~~11 -- Correlation applicability conditions & 368\\
~~~12 -- Advantages of correlation & 368\\
~~~13 -- Comparison of two correlation coefficients & 372\\
~~~14 -- Correlation ratio & 375\\
~~~15 -- Determination and alienation coefficients & 376\\
~~~16 -- Correlation of the number of cases & 376\\
~~~17 -- Location correlation & 377\\
~~~18 -- Conclusions & 378\\[3pt]
23 -- Statistical adjustment & 379\\
~~~1 -- Adjustment concept & 380\\
~~~2 -- Adjustment methods & 380\\
~~~3 -- Linear adjustment & 388\\
~~~4 -- Curve fitting & 393\\
~~~5 -- Adjustment to theoretical distributions & 400\\[3pt]
24 -- Checking the normality of a series & 406\\
~~~1 -- Graphic appreciation & 407\\
~~~2 -- Cumulative frequencies & 408\\
~~~3 -- Straight from \textsc{Henri} & 409\\
~~~4 -- Significance of kurtosis and asymmetry measurements & 411\\
~~~5 -- Fitting to a normal distribution & 415\\
~~~6 -- The $\ll$Probits$\gg$ & 420\\[3pt]
\end{tabular*}

\begin{tabular*}{0.85\textwidth}{@{\extracolsep{\fill}} lr } 
25 -- $\ll$Probits$\gg$ Method & 420\\ 
~~~1 -- Principles of the method & 420\\
~~~2 -- Checking the normality of a distribution & 424\\
~~~3 -- Calculation of lethal dose & 425\\[3pt]
26 -- Chronological series & 433\\
~~~1 -- Definition & 433\\
~~~2 -- Movements of a chronological series & 434\\
~~~3 -- Guidance for studying  series & 434\\
~~~4 -- Data preparation & 435\\
~~~5 -- Graphical representation & 436\\
~~~6 -- Analysis of time series & 440\\
~~~7 -- Study of the general trend & 441\\
~~~8 -- Seasonal trend & 444\\
~~~9 -- Cyclic movements & 453\\
~~~10 -- Comparison of time series & 453\\[3pt]
27 -- Index numbers & 456\\
~~~1 -- Concept & 457\\
~~~2 -- Types of indexes & 458\\
~~~3 -- Conditions for displaying an index number & 462\\
~~~4 -- Problem of creating an index & 462\\
~~~5 -- Random errors in index calculations & 464\\[3pt]
\multicolumn{2}{c}{\textsc{Second part}} \\[3pt]
Basic notions of biostatistics & 465\\
~~~0 -- Settings & 465\\
~~~1 -- Population & 466\\
~~~~~~1.1 -- Population status & 466\\
~~~~~~1.2 -- Population movement & 478\\
~~~~~~1.2.1 -- Birth & 478\\
~~~~~~1.2.2 -- Mortality & 486\\
~~~~~~1.2.3 -- Demographic balance & 497\\[3pt]
\multicolumn{2}{c}{\textsc{Third part}} \\[3pt]
Annex I -- Statistical terminology & 500\\
Annex II -- Statistical symbols & 501\\
Bibliography & 503
\end{tabular*}}

\newpage

\section*{Appendix 3}
Table of Contents of  {\it Introdução à Estatística}  \cite{DiasAgudo1961} [{\it Introduction to Statistics}], notes for the course on General Mathematics for the degrees in Biological Sciences, Geological Sciences, and Adjunct Professors, based on the lectures of Professor Dias Agudo, compiled by Moreira Campos / Waldemar Nunes, Lisbon 1961/62
\bigskip

{\footnotesize
\begin{tabular*}{0.85\textwidth}{@{\extracolsep{\fill}} lr } 
\multicolumn{2}{l}{Chapter I -- Fundamental Concepts of Probability Theory}\\[2pt]
~~~~~~1.1 -- Introduction & \\
~~~~~~1.2 -- A fundamental principle& \\
~~~~~~1.3 -- ``Arrangements and Combinations''& \\
~~~~~~1.4 -- Event space and probability space& \\
~~~~~~1.5 -- Probabilities of the sum and product of events.& \\
~~~~~~~~~~~~~Conditional probabilities.& \\
~~~~~~~~~~~~~Incompatible and independent events& \\
~~~~~~1.6 -- Problem with repeated trials.& \\
~~~~~~~~~~~~~Binomial formula& \\[3pt]
\multicolumn{2}{l}{Chapter II -- Random Variables and Distribution Functions}\\[2pt]
~~~~~~2.1 -- Casual variables& \\
~~~~~~2.2 -- Distribution function& \\
~~~~~~2.3 -- Classification of distributions& \\
~~~~~~2.4 -- Functions of casual variables& \\
~~~~~~2.5 -- Percentiles of a distribution& \\[3pt]
\multicolumn{2}{l}{Chapter III -- Discrete Distributions}\\[2pt]
~~~~~~3.1 -- Probability function; elementary probabilities& \\
~~~~~~3.2 -- Mean value& \\
~~~~~~3.3 -- Moments& \\
~~~~~~3.4 -- Important examples of discrete distributions& \\
~~~~~~~~~~~~~Poisson distribution& \\
~~~~~~~~~~~~~Binomial Distribution& \\[3pt]
\multicolumn{2}{l}{Chapter IV -- Continuous Distributions}\\[2pt]
~~~~~~4.1 -- Probability density& \\
~~~~~~4.2 -- Mean value. Moments. Median& \\
~~~~~~4.3 -- Important example of continuous distribution: Normal distribution& \\
~~~~~~~~~~~~~4.3.1 -- Reduced normal distribution& \\
~~~~~~~~~~~~~4.3.2 -- General normal distribution& \\[3pt]
\multicolumn{2}{l}{Chapter V -- Sums of Random Variables}\\[2pt]
~~~~~~5.1 -- Mean value and variance of the sum& \\
~~~~~~5.2 -- Distribution of the sum of casual variables& \\
~~~~~~5.3 -- Normal distribution as an approximation of binomials& \\[3pt]
\multicolumn{2}{l}{Chapter VI - Sampling}\\[2pt]
~~~~~~~~~~~~~Data Presentation and Description& \\
~~~~~~6.1 -- Introduction& \\
~~~~~~6.2 -- Frequency distribution; histograms; data classification& \\
~~~~~~6.3 -- Location  and dispersion measurements& \\[3pt]
\end{tabular*}}

{\footnotesize
\begin{tabular*}{0.85\textwidth}{@{\extracolsep{\fill}} lr } 
\multicolumn{2}{l}{Chapter VII -- Parameter Estimation}\\[2pt]
~~~~~~7.1 -- Introduction& \\
~~~~~~7.2 -- Mean value estimation& \\
~~~~~~7.3 -- Variance estimation& \\[3pt]
\multicolumn{2}{l}{Chapter VIII -- Hypothesis Tests}\\[2pt]
~~~~~~8.1 -- Introduction& \\
~~~~~~8.2 -- Hypothesis tests on the mean value& \\
~~~~~~8.3 -- Hypothesis tests on the type of distribution& \\
~~~~~~8.4 -- Comparison of two populations& \\
\end{tabular*}}

\subsection*{Acknowledgments} 
This work is partially financed by national funds through FCT --- Fundação para a Ci\^{e}ncia e a Tecnologia under the project UIDB/00006/2020 \url{https://doi.org/10.54499/UIDB/00006/2020}.
 
\noindent The authors are thankful to Prof.\ Ant\'{o}nia Amaral Turkman, Maria de F\'{a}tima Brilhante and Sandra Mendon\c{c}a for useful observations on the original draft. The authors also extend their gratitude to the reviewers for their stimulating comments, corrections and informations that helped to improve the paper.

{\small\bibliography{cimart}}
% Please, do not change the above line and do not insert your references
% into this file.  Instead, insert your references into the cimart.bib file.
% See cimart.bib for further instructions.

\EditInfo{July 30, 2024}{October 4, 2024}{Ana Cristina Moreira Freitas, Oliveira E Silva Diogo, Ivan Kaygorodov and Carlos Florentino}
\end{document}